\documentclass[sigconf]{acmart}
\usepackage{url}
\usepackage{multirow}
\usepackage{graphics}
\usepackage{enumitem}
\usepackage{color, colortbl}
\usepackage{lipsum}
\usepackage{tabularx,booktabs}
\usepackage[section]{placeins}
\usepackage{graphicx}
\usepackage{comment}
\usepackage[para]{footmisc}
\usepackage{rotating}
\usepackage{tablefootnote}
\restylefloat{table}
\usepackage{breakurl}
\usepackage[nobiblatex]{xurl}
\definecolor{lightgray}{rgb}{0.83, 0.83, 0.83}
\usepackage{caption}
\usepackage{subcaption}
\usepackage{mwe}    

\DeclareCaptionLabelFormat{simplab}{Figure~#2}
\DeclareCaptionFormat{nocap}{}
\DeclareCaptionSubType*{figure}

\makeatletter
\def\@copyrightspace{\relax}
\makeatother

\makeatletter
\newcommand\notsotiny{\@setfontsize\notsotiny{5.5}{6.5}}
\makeatother

\usepackage{caption}
\copyrightyear{2023}
\acmYear{2023}
\setcopyright{acmlicensed}\acmConference[WiSec '23]{Proceedings of the 16th ACM Conference on Security and Privacy in Wireless and Mobile Networks}{May 29-June 1, 2023}{Guildford, United Kingdom}
\acmBooktitle{Proceedings of the 16th ACM Conference on Security and Privacy in Wireless and Mobile Networks (WiSec '23), May 29-June 1, 2023, Guildford, United Kingdom}
\acmPrice{15.00}
\acmDOI{10.1145/3558482.3590173}
\acmISBN{978-1-4503-9859-6/23/05}

\begin{document}



\date{}

\title{Commercial Anti-Smishing Tools and Their Comparative Effectiveness Against Modern Threats}

\author{Daniel Timko}
\affiliation{%
  \institution{California State University San Marcos}
   \city{San Marcos}
   \state{CA}
   \country{USA}}
\email{timko002@csusm.edu}

\author{Muhammad Lutfor Rahman}
\affiliation{%
  \institution{California State University San Marcos}
   \city{San Marcos}
   \state{CA}
   \country{USA}}
\email{mlrahman@csusm.edu}

\thispagestyle{empty}

\begin{abstract}
Smishing, also known as SMS phishing, is a type of fraudulent communication in which an attacker disguises SMS communications to deceive a target into providing their sensitive data. Smishing attacks use a variety of tactics; however, they have a similar goal of stealing money or personally identifying information (PII) from a victim. In response to these attacks, a wide variety of anti-smishing tools have been developed to block or filter these communications. Despite this, the number of phishing attacks continue to rise.
In this paper, we developed a test bed for measuring the effectiveness of popular anti-smishing tools against fresh smishing attacks. 
To collect fresh smishing data, we introduce Smishtank.com, a collaborative online resource for reporting and collecting smishing data sets. The SMS messages were validated by a security expert and an in-depth qualitative analysis was performed on the collected messages to provide further insights. 
To compare tool effectiveness, we experimented with 20 smishing and benign messages across 3 key segments of the SMS messaging delivery ecosystem. Our results revealed significant room for improvement in all 3 areas against our smishing set. Most anti-phishing apps and bulk messaging services didn't filter smishing messages beyond the carrier blocking. The 2 apps that blocked the most smish also blocked 85-100\% of benign messages. Finally, while carriers did not block any benign messages, they were only able to reach a 25-35\% blocking rate for smishing messages.
Our work provides insights into the performance of anti-smishing tools and the roles they play in the message blocking process.
This paper would enable the research community and industry to be better informed on the current state of anti-smishing technology on the SMS platform. 
\end{abstract}

\maketitle
\section{Introduction}
\hspace{0.4cm} In 2021, phishing was responsible for 90\% of data breaches in the US~\cite{cisco}, and smishing attacks, in particular, have been reported by 74\% of organizations~\cite{proofpoint}. 
Due to the increase of fraudulent communications~\cite{fbi}, smishing is gaining a lot of attention by cybersecurity experts. These attacks take advantage of several legitimate services to perform and conceal their activities, thereby making smishing difficult to combat. 
For example, URL shorteners like bitly~\cite{bitly} can be used to hide the real destination that a link will forward the user to. As with SMS advertising, an attacker can use or purchase numerous cell phone numbers, known as cell phone leads, and distribute smishing messages to them through a wide array of bulk messaging services operating as External Short Message Entities (ESME).
 
 To counter the rise of phishing messages, an array of anti-smishing tools have been created to block and filter SMS. These anti-smishing tools
and built-in anti-smishing security provided by mobile carriers, constitute the technical means for preventing smishing on mobile devices. Each tool can use a variety of techniques to detect phishing messages, where some are publicly known and others unknown. On the legal end of matters, laws have been passed to regulate the sending of unsolicited messages~\cite{TCPA}. However, this has not stopped the recent growth of smishing campaigns~\cite{smishincrease}.

There are several studies which compare the effectiveness of phishing tools and techniques~\cite{comparativeanalysisphishtools,zhang2007phinding,surveyofphishtools,impactofantiphishingtools}. These studies have provided important insights for end-users, researchers and developers on the performance of commercial anti-phishing technology. However, to the best of our knowledge, no study has been done on the efficacy of commercial anti-smishing tools. With the increase in smishing attacks, it is important to determine whether these tools effectively serve their intended purpose of blocking smishing attacks. The detection and prevention of smishing campaigns is multifaceted. Messages pass through several filters on their path to a target's inbox. To determine the state of commercial anti-smishing technology we can isolate the effect each filter plays on the messaging process through experimental testing. This analysis is important to tackling the growing trend of smishing messages as it will guide us to where the gaps in smishing detection and prevention lie. With these answers, we can derive information on where future smishing research should be focused and make recommendations on how the tools could be improved. 
Fresh and unvalidated smishing communications will be used to test these tools more accurately against modern threats~\cite{zhang2007phinding}. 
This is because, as time goes on, phishing messages and mitigation approaches change. Furthermore, some messages will find their way onto blacklists, but by then, damage will have already been done. Accordingly, the objective for this research is to study the effectiveness of tools against modern smishing attacks. For this work, we treated the smishing techniques like a black box. Instead of analyzing how this technology works, we focus on how well they work.
 
A secondary task to this research is to provide a publicly available source of fresh phishing messages for testing purposes. Researchers rely on publicly available data sets for phishing research~\cite{smsbulkmessaging}; however, there is a lack of publicly accessible fresh smishing feeds. In several recent studies, researchers have relied on older publicly available data sets instead of collecting newer ones~\cite{smsclasstopicmodeling}. While some fresh smish feeds exist, they are maintained by private SMS communication providers and anti-smishing tools for their blacklists. These types of private feeds require permission to access. Additionally, even if it is possible to 
gain access to the private blacklist feed, by the time you can access the smishing data, they can no longer be "considered fresh" after they reach their blacklists and have been processed by their classification algorithms. Despite the best efforts of these private smishing detection systems, end users still receive phishing messages.
Therefore, researchers will need access to fresh messages that break through those detection systems to study and develop new approaches. Some raw SMS communication feeds are available online through public SMS gateways, which could be used to gather smishing SMS corpus. However, abusive messages, such as smishing communication, comprise only a small fraction of the messages and lack filtering~\cite{smspublicgateways}. Alternatively, there are publicly available phishing feeds for website URLs and email phishing, but they have different phishing identifiers and composition from smish. Furthermore, they cannot be used interchangeably for smishing detection~\cite{phishtank,apwg}. Mobile solutions are needed which can provide similar online phishing databases for user communities~\cite{avoidphishbait}. In this project, we will use crowdsourcing to gather the smish through the Smishtank~\footnote{https://smishtank.com/} site and publicly display the submissions through a feed.

\textbf{Our Contributions:} With this work we provide several novel contributions towards the mitigation of smishing attacks. More specifically, we produced the first comparative study of anti-smishing technology across bulk messaging services, carriers and anti-smishing apps, regarding the accuracy of smishing detection and filtering. \textbf{This work made the following contributions:}

\begin{enumerate}[leftmargin=*]
   \item 
   We deployed a community driven resource for collecting user submitted phishing messages. 
   \item We breakdown and analyze our collected messages across several categories to provide insights into the smishing attacks.
   \item We explored the blocking rate of 5 carriers, 5 bulk messaging services, and 10 anti-smishing apps against smishing attacks. 
   \item We conducted a three-part experimental research design to compare and analyze the performance of modern anti-smishing tools against new attacks.
 \end{enumerate}
\section{Related Works}
\label{sec:relatedworks}
\paragraph{Gathering Datasets}
A critical step to providing a detailed comparative analysis of anti-smishing tools is gathering SMS messages for testing. To test these anti-smishing tools against the latest attacks, the work by Zhang et al. suggests that we use the freshest samples we can find~\cite{zhang2007phinding}. We define the freshness of a smish as a measure of how recently it has been sent to the device. 
In terms of smish, this is a difficult prospect due to limited availability of public data sets~\cite{smsspamfilteringcollection,Mishra2021DSmishSMSAST}. Research by T. Almeida on SMS spam filtering also encounters similar issues based on the lack of available public SMS data sets. The research describes how lack of available data hinders research on SMS filtering classification~\cite{smsspamfilteringcollection}.
To gather smishing data sets, researchers have relied on many novel approaches to collect smishing data~\cite{honeypot, smspublicgateways}.
One method for collecting these phishing messages is through community submissions.
 S. Baadel presents the importance of online user communities in the process of gathering fresh data~\cite{avoidphishbait}. 
In terms of email and website phishing, phishing feeds exist such such as Phishtank~\cite{phishtank} and APWG~\cite{apwg}. 
They provide a community submission and validation system of phishing URLs for researchers and the public through their open API. There have also been archive efforts to store older phishing URLs from these sources by using projects such as Phishmonger~\cite{phishmonger}. 
 
\paragraph{Smishing Detection and Filtering}

\hspace{0.4cm}There are several aspects to detecting and filtering phishing messages in anti-smishing tools. Previous work has simplified the taxonomy of phishing detection methods~\cite{taxonomy}, as well as described the process of phishing detection and filtering mechanisms~\cite{smsphishingandmitigation}. The most commonly used approaches for smishing detection are list-based and content-based filters. List-based filtering such as blacklists, whitelists and greylists are used with lookup tables. 
Blacklists are the most common line-based filtering mechanism which are used against phishing, but they have many known drawbacks~\cite{Mishra2021DSmishSMSAST, taxonomy, JAIN2018617}. A. Oest's work analyzes the effectiveness of blacklists against URLs used in phishing attacks~\cite{blacklisteffectiveness}. Notably, the use of URL shortening services are a problem for blacklists, and are common in smishing attacks. Content-based filtering, including heuristics and machine learning approaches are also popular for phishing detection and classification.
There are numerous works utilizing machine learning models for phishing detections~\cite{phishnet,phishbox,JAIN2018617}. A unique feature of content-based filtering over list-based filtering is in their ability to detect zero-day smishing attacks~\cite{JAIN2018617}.
 
\paragraph{Bulk messaging, ESME and Gateways}
 \hspace{0.4cm} In the past few years much work has been done to address the security concerns of bulk texting through SMS gateways. Laws, such as Telephone Consumer Protection Act(TCPA) in the US, state that advertisers must gain explicit consent before texting consumers. However, despite efforts by legitimate bulk messaging services and legal remedies, gateways are being used to deliver malicious SMS messages~\cite{smspublicgateways}. While legitimate ESMEs adhere to legal guidelines and restrictions, there has been news of illegitimate ESMEs and gateways which market themselves as "crime-friendly"~\cite{krebs_2021}. Therefore, it's important to define the role ESME have in curbing smishing attacks as they can be used to distribute mass smishing campaigns through bulk texting services. B. Reaves provides a detailed overview of the role SMS gateways play in the security infrastructure~\cite{smspublicgateways}. Within the overview, it is mentioned that these ESMEs provide the interface for the bulk texting services.

\paragraph{Phishing Tool Analysis}
\hspace{0.4cm}Several studies have been conducted on the comparative detection rates of anti-phishing tools~\cite{surveyofphishtools,zhang2007phinding,comparativeanalysisphishtools}. Anti-smishing tools are integrated into the applications that we use for texting, as well as the browsers that we use to open links from our SMS. Zhang et al. presents a comparative analysis of anti-phishing toolbars~\cite{zhang2007phinding}. The researchers present a structure for evaluating phishing tools based on freshness and catch-rate. Moreover, the study contains an analysis of how phishing attacks bypass detection algorithms. The work by Zeydan et al. mentions that the implemention of anti-phishing tools occurs at multiple levels~\cite{surveyofphishtools}. It is integrated into the software that we use as well as server and client-side solutions. 
Work by Chorghe et al. on anti-phishing techniques for mobile browsers observed that mobile devices are more susceptible to zero-day phishing attacks and notes the lack of research into anti-phishing tools on smartphones~\cite{surveyantiphishtechniques}.
\section{Background}
\label{sec:background}
\begin{figure}[htbp]
    \centering
        \includegraphics[width=.49\textwidth]{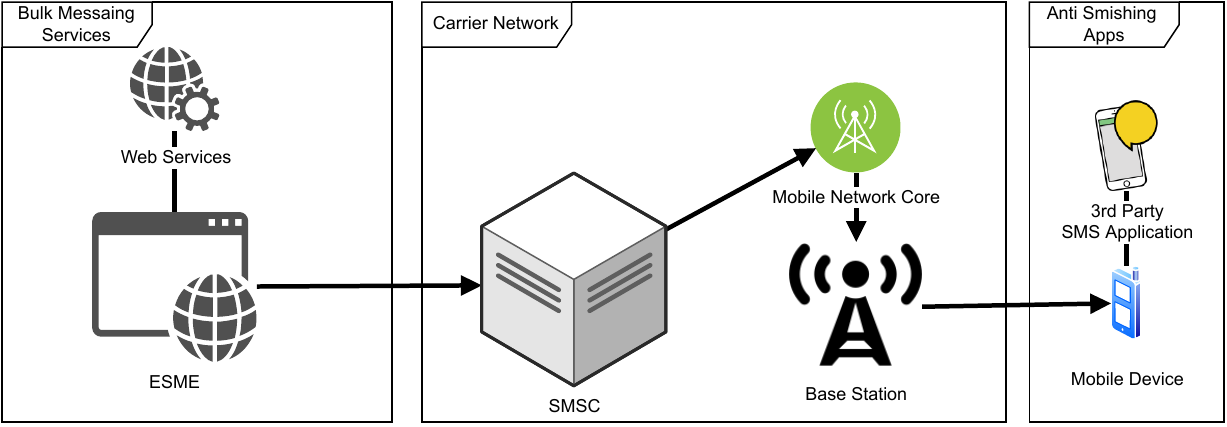}
    \caption{The mobile messaging ecosystem separated into different components.}
    \label{fig:BulkMessagingProcess}
    \vspace{-3mm}
\end{figure}
We separate the messaging ecosystem into 3 broadly defined sections corresponding to the bulk messaging service, mobile carrier, and 3rd party apps. Each of these defined sections can be found in Figure~\ref{fig:BulkMessagingProcess}. Previous work by B. Reaves analyzes the modern SMS messaging ecosystem within cellular networks~\cite{smsbulkmessaging}. In their analysis, the messaging process from ESME to target phone is described. Initially, an ESME which provides Application-To-Person(A2P) web services is used to send the message data. ESMEs then deliver this messaging data to Short Message Service Centers(SMSC), which reside within mobile networks. SMSC, in turn, store and forward messaging data over the core network to base stations, which handle communication with individual mobile devices. This message delivery process is complex and the message traffic can involve multiple SMSC and other entities~\cite{IDACHABA201496,sendingoutsms}, each potentially implementing their own SMS firewall~\cite{smsgatewayfirewall}.
The implementations of carrier network security or their messaging data are not typically public. Additionally, breaking down the contribution of each particular entity inside the carrier network space would provide valuable insights, but it is outside the scope of this project. Instead, we bundle filtering entities inside the carrier network space and analyze them as they are delivered to the User Equipment that constitute our tested mobile devices. In our research, we analyzed the routing of bulk SMS messages sent through ESMSE to a target phone over a carrier network.
\section{Methodology}
\label{sec:methodology}

\hspace{0.4cm} Since the purpose of this experiment is to compare the efficacy of various popular anti-smishing tools, we isolated 3 sections of the messaging process in which the smish would be filtered or blocked. The message filtering techniques of anti-smishing tools can potentially utilize a wide array of text and non-text based factors in their classification. In our experimentation, we aim to maintain as realistic of a phishing scenario as possible. A visualization of the entire message sending procedure, including the message sending paths, is illustrated in Figure~\ref{fig:testingmethodology}. Messages were sent in 2 sets of 20 messages, the first set benign and the second smishing, to each target. 
These sets, when combined, contain all messages in our message list. Similar to related works~\cite{impactofantiphishingtools,phishbox}, we used a balanced set between smishing and benign.

Our Analysis plan consists of 3 different chi square tests. 
Additionally, when at least one cell count is less than 5 we instead use Fisher's exact test. As a result, the comparative \textit{smish hit rate} and \textit{benign strike rate} of messages sent through these sections are covered. Here, the \textit{smish hit rate} is the percentage of smish caught, and the \textit{benign strike rate} is the percentage of benign messages blocked. First, we analyze how bulk messaging services respond to smishing campaigns by blocking suspicious messages from being sent. Next, we analyze how carriers respond to receiving smishing messages. Carriers can use built-in message filtering or refuse to deliver the message to their recipients through carrier owned networks. Likewise, a carrier can also report back to the bulk texting service that the messages have been blocked and so has the fraudulent scammer. Finally, we analyze how third-party anti-smishing applications detect and block smishing messages. These apps work on top of the carriers' blocking capacity to filter received messages and separate them into a spam folder. Such apps have a wide range of features; however, for the purpose of this work we only look at apps that state they filter SMS messages. A brief description of each independent and dependent variable discussed in this work can be found in Table~\ref{tab:variable-table}.
Among our carriers, we randomly selected T-Mobile to serve as the target carrier service used in part 1 and 3 of our testing (Table~\ref{tab:carrier-table}).

This research involved a four-step workflow process. First, we collected phishing messages from the community and selected the 50 most recent validated phishing messages for testing. Additionally, we collected 50 ham SMS messages from the UCI machine learning SMS spam collection~\cite{UCICollection}. Second, we processed the data by compiling it into a list of fresh unvalidated phishing messages. We then worked alongside a security expert to validate our message set before sending it out. Third, we created a test bed by developing an app to integrate with available API services. We also set up bulk messaging service accounts for all tools and mobile devices used in this testing. Finally, we performed three tests over the course of three days from May 19, 2022, to May 21, 2022, to send over 1000 combined text messages. In each test, we isolated the effect of one factor on the message blocking process.

\subsection{SMS Collection}

\hspace{0.4cm} In this work we use community submissions to collect our smishing SMS messages. While similar resources exist to obtain large numbers of phishing URLs, few exist to obtain public phishing messages. 
Additionally, we cannot simply use URLs to create a complete picture of smishing, as it is only one part of smishing detection on mobile. Smishing poses unique detection challenges that require a diverse data set. The sender ID, for instance, can be a phone number, short code, email address or a brand name, and due to the open nature of SMS API and gateways, these fields can be spoofed. 
Therefore, we have developed our own community-based smishing submission site to collect phishing messages through user submissions located at {\url{https://smishtank.com}}.
Visitors to the site can submit phishing messages into our database.

All phishing messages used in this research were collected over the course of 2 months. This ensures that the messages are as "fresh" as possible to test our tools against newer threats. 
Messages which have potentially already passed through a filtering system may end up on the block list of one of the tools we use in testing. This limits our data set to messages we collect ourselves.
Additionally, due to the nature of public access of community submissions, we will not include messages that contain personally identifying information(PII)
. As part of our ethical considerations, we remove messages with PII from the public submission list.

Through our collection of phishing messages from smishtank, we were able to gather 75 total message submissions. Submissions which were identified as duplicates, jokes or memes were removed from the collection and 10 additional smishing messages were added from recent social media posts on Redditt (@r/scams), and Twitter. From the resulting data set we submitted 55 messages to a security professional for proper smishing validation. The security professional has more than 9 years of phishing-related research experience. The validation of the smishing messages was based on a combination of factors including the message contents, links, and similarity to phishing scenarios. After validation, the resulting set available for testing included all 55 messages. From this pool of available smish, we randomly selected 20 smishing messages. 
A full list of the messages obtained and used in this research can be found on the smishtank website\footnote{https://smishtank.com/wisec23data \label{smishtank}}.
Next, we selected benign messages from the UCI Machine Learning Repository's SMS spam collection~\cite{UCICollection}. This data set contains both Ham and Spam classified SMS messages. 
We randomly selected 20 Ham messages in this data set. The data set contains various typical SMS messaging features, such as emoji text. We utilize only smishing and ham messages to ensure that blocking will be due to smishing and not spam.

\subsection{Data Characterization}
\label{subsec:datacharacterization}

\hspace{0.4cm}To understand the messages used in this work we performed a qualitative analyses on our SMS collection. This allows us to provide a broader depth of analysis by studying the characteristics of the smishing messages. An additional factor that we considered was the difference between spam and smishing. Foozy et al. clearly distinguished between spam and smishing~\cite{smishingfromspam}. These smishing messages are distinct from a spam collection in that they purposely attempt to defraud their target. While spam is sometimes used as a catchall for illegitimate communications, we define it as unwanted marketing or advertising. In utilizing only smish, our messages contain elements that are deceptive. Based on our original set of 55 validated smishing messages, we broke down each message into several categories based on their Named Entities, and Subcategories. Similarly, we perform this analysis on the 20 smish selected for the experiment. Of the 55 messages collected, 46 contained URLs (18 for the selected set), which we further analyze for URL types and squatting techniques. The resulting table can be found in Table~\ref{tab:data-characterization}. When possible, we reference examples from our message list located on smishtank\textsuperscript{\ref{smishtank}}.

\paragraph{Squatting Techniques}
The purpose of squatting is to trick unsuspected users into misidentifying a malicious domain for a well-known one. These Squatting techniques have been broken down into 5 categories by Quinkert et al.~\cite{bethephisher} which we used to identify Squatting techniques of our smishing set. 
 Surprisingly, more than half of our domains 60.87\% did not employ any squatting techniques at all. Instead, they utilized links made up of random or semi-relevant words to the message subject matter. Alternatively, we observed that some messages contained multiple squatting techniques. The most common squatting technique in our messages was Combosquatting. Combosquatting adds additional terms to a real domain to form authentic looking urls (e.g., verifywellsfargo.ga).
Wrong TLD use different top-level domains from the authentic site (e.g., jf245-fedex.me). Typosquatting domains contain spelling errors that differentiate it from the real domain (e.g., uvusps.com). Subdomain Usage refer to domains that contain additional subdomain labels (e.g., usps.informol.com). Finally, Homograph squatting use look-alike characters to trick users visually (e.g., cit1us3rinf0.com). 
 
\paragraph{URL Types} Work by Oest et al. \cite{insidephishersmind} presents a classification scheme for phishing URLs using 5 separate types of phishing URL classifications. We apply these classifications to each final URL destination in our message set. In addition to the URL types, we also observed different URL obfuscation tactics using URL shorteners (6/46) and redirects (14/46) and resolved them to their destination for this analysis. More than half of the final message URLs (71.74\%) contained deceptive top-level domains. These messages contained brand names or deceptive keywords in the domain which were tailored to the phishing scenario. A random domains with deceptive path content (e.g., stopsoriasis.co.il/-/verify?fifthird) contains a deceptive terms after the top-level domain. Deceptive Subdomains (e.g., www.secre.citi.us.accesauth.online) contains deceptive subdomain keywords. The unintelligible URLs (e.g., f2gpy.info/RzNKEwsZve) contain no keywords in either the path or domain related to the phishing topic. Finally, None of the messages we received contained IP addresses as the hostname.
 
\paragraph{Named Entities} Phishing messages often include recognizable names or entities. The purpose of including these names is to impersonate entities that a potential target may trust, in order to get them to perform an action. Of the 55 messages, we found that 41.82\% did not mention any well known entities. The most common named entities were banks (e.g., Citizens Bank). Postal service entities (e.g., Fedex), are shipping services. Individual entities use personal greetings from random numbers, typically with selfies. Social Media entities (e.g., Snapchat), are often used to invite users to private chats. Some entity types which only appeared once in our message set(e.g., Costco), were labeled Other. 
 
\paragraph{Subcategories} Profiling smishing messages can help us to identify patterns and to measure the types of strategies used by malicious individuals. We labeled our smishing messages into 10 subcategories using categories similar to previous work~\cite{cluesintwitter,smsbulkmessaging}. We found that Account Alert messages comprised of the plurality of the messages received. These messages notify targets that their account has been compromised (e.g., M9).
The Prize/Contest category correspond to messages that offer free prizes, or state that you have won a contest (e.g., M1). 
Delivery type scams (e.g., M6) use a notification of an undelivered package and typically request PII to claim the package. Payday Loan/Credit scams (e.g., M14) entice their target with claims of easy access to loans or credit. Wrong Number/Romance Scams, pretend to mistake the target for someone they know and often use flirtation or contain pictures of attractive women. 
Job advertisements (e.g., M13) invite the target to apply for a job. Link only messages (e.g., M5) contain links and no additional content. Finance/Crypto related scams (e.g., M3) offer financial or crypto trading, often with promises of easy money. Lawsuits/Settlement scams (e.g., M11) discuss either a potential lawsuit or claim that there is an available cash settlement. Lastly, advertisement contained offers to sell or buy a product.

\paragraph{VirusTotal and Domain History}
\begin{table}[h]
\scriptsize
\centering
\vspace{-5mm}
\begin{tabular}{@{}llllllll@{}}
\toprule
Type & \#N & VT\textgreater{}1 & VT\textgreater{}5 & VT\textgreater{}10 & VT-M & VT-MW & VT-P \\ \midrule
URLS(ALL) & 46 & 43.48\% & 15.22\% & 8.70\% & 28.26\% & 19.57\% & 26.09\% \\
Domains(ALL) & 44 & 25.00\% & 11.36\% & 9.09\% & 20.45\% & 6.82\% & 20.45\% \\ \midrule
URLS(Selected) & 18 & 72.22\% & 11.11\% & 11.11\% & 50.00\% & 22.22\% & 38.89\% \\
Domains(Selected) & 17 & 29.41\% & 5.88\% & 5.88\% & 23.53\% & 11.76\% & 23.53\% \\ \bottomrule
\end{tabular}
\caption{Virus Total reports for all smishing messages and the messages randomly selected for the experiment. }
\label{tab:virus-totals}
\end{table}
 In order to better understand what kind of threat these messages posed, we analyzed them through VirusTotal~\cite{virustotal}. The analysis was performed 7 months after the initial experiment, giving time for messages to be reported and find their way onto vendor blacklists. VirusTotal collaborates with over 70 vendors to provide detailed results from scanning files, URLs and domains. Using this service, we looked at the status and domain update timeline for the messages set. In table~\ref{tab:virus-totals} we note the status of the URLs and their domains, as well as the subsequent number of vendors that rated the URLs as malicious. Interestingly, even after 7 months, only a fraction of messages that were identified as malicious were labeled by more than 5 vendors. We also observed that the status of messages changed through redirects and shorteners. By analyzing the final URL instead of original URL, the status of 6 of our redirected URLs changed from malicious to clean. Potentially, the malicious sender could reuse many of their vendor blacklisted URLs by just applying a new URL shortener or redirection path. Additionally, among the messages marked as clean by vendors, we were able to find examples~\cite{onlinethreatalerts} of online communities that discussed those messages as scams. 

Next, we look at the Domain history of the websites referenced in our smishing messages. In terms of creation time, we can positively confirm that (30/46) of the domains were created within 1-3 months of us receiving the messages. When considering last update time, that number jumps to (36/46). The domain timeline of (3/46) messages could not be confirmed due to their unavailability. Another (5/46) messages utilized authentic domains to deliver their smishing messages, so their older creation date is unsurprising. These messages used either private groups on social media sites, or web API to begin phone calls. Next, we witnessed 2 older domains (20+ years old) that were being used to host malicious sites. Surprisingly, when we investigated the domain owner we found that they had been involved in civil suits before for hosting malicious content~\cite{vitalworks}. Finally, out of the 46 collected URLs, only one has been updated since our experiment.

\subsection{Participants}

\hspace{0.4cm} In order to create more realistic testing conditions we recruited participants to test our message set against their mobile devices. We understand that including participants introduces risk, but we took mitigation steps to minimize this harm and consider that the benefits outweigh the risks. This research was done with approval from our university’s Institutional Review Board (IRB). Participants were recruited primarily through word-of-mouth and our university email. We recruited participants based on their phones' carrier and platform to match our testing requirements. More precisely, the requirements consisted of 10 T-Mobile devices, 5 Android, 5 iOS, and one device for each carrier service. In total, 14 participants were recruited for this experiment. One T-Mobile phone served as the recipient to test the carrier service and anti-smishing app.

To uphold testing ethics and mitigate harm, we accompanied our participants in a zoom call to instruct them throughout the full testing process. First, we verified with the participants if they have any messaging applications or message filtering software installed before the testing begins. We guided participants on how to set the built in messaging app to the default messaging service for the duration of the test, and instructed them on how to set it back afterwards. In part 3 of our experiment, we instead requested that participant install an anti-smishing app, and set that app as the default messaging service. This process allows the messaging app to take control of message intents sent to the device and is a necessary step to the filtering process. A limitation in this setup is that for ethical reasons we do not have physical access to the participant's phone. Prior to sending, we inform participants of the number and sender of messages. 

Once the participants are ready, we deliver messages to their device through the test bed. After messages are sent we instruct participants on how to safely screenshot the conversation with our bulk messenger and deliver those screenshots back to us through email or text. Afterwards, for safety reasons we request that they delete the conversation from their device. In this aspect, we limit the number of messages each participant receives to 40 to reduce burden and room for error when sending and taking screenshots of the messages. Subsequently, we used these screenshots provided by the participants to verify the messaging tests. The screenshots contain the text of the messages, when we sent them as well as the phone number used to deliver the message. No PII or participant phone numbers are included in the screenshots. This testing process took between 5 to 10 minutes per person. Finally, we anticipated privacy concerns with participants providing their contact information for this study. Participant's consent and information was submitted through our Qualtrics form and was stored securely on a cloud drive. Similarly, message screenshots were stored securely with the participant data. This data is encrypted to prevent unauthorized access. Only the investigators have access to this data. This information will be stored for no longer than 6 months.

\subsection{Tool Testing Process}

\hspace{0.4cm} The first step in our tool testing process was to build our test bed. While these bulk messaging services allow you to send individual messages through through their websites, API access is required to automate the process for a large number of messages and targets. We set up a local server with an app front-end which connected, through our local server, to the API endpoints hosted by the bulk messaging services. 
Our app allowed us to select one of the bulk texting services and then read a list of numbers and messages from a file to send out a group of messages at once. After sending, we observed all the JSON callbacks and individually verified the messages that were received on the intended target device. Due to testing requirements, we recruited participants to allow us to send messages to their phones and then relay the messaging list with screenshots back to us for manual verification. 
To send the messages required for this research, we also purchased subscription or top-up option from each messaging service and a toll-free number if one is not provided by the service for testing. Testing was done from toll-free numbers to target phones with local numbers. Typically, a texting service strictly scrutinize message sending for trial or free accounts. Additionally, while a high volume A2P local number would have a higher message delivery rate, we primarily used toll-free numbers to maintain consistent results~\cite{localnumberbetter}. 

The tools used for this research include third-party anti-smishing apps and the built-in anti-smishing services provided by mobile carriers. We were able to test all apps, carriers, and bulk messaging services on the same set of messages to identify specific blocking tendencies. All tests were performed within a 3-day period.

This study only considers successful filtering of phishing messages as positive identification. To this end, messages filtered into a spam folder, identified as phishing or removed from the inbox are considered positive identifications. Warnings are excluded from the analysis because they can introduce bias into the algorithms. The SMS messages are not edited to preserve their original form. To avoid interruptions by bulk messaging services, target phones are changed between submissions and if messaging is interrupted, messages are resent with a new account and target phone.

\subsection{Overview of the Bulk Messaging Services}
\hspace{0.4cm} We have selected 5 ESME which constitutes our bulk messaging services to test our phishing messages. These ESME services can provide access to phone numbers for their texting services, which can send out the messages untethered to any SMS carrier. It's observed that these bulk messaging services can offer either local or toll-free numbers, often accompanied by a monthly fee. While each tool may have a different policy for handling its service, they all state that they follow CTIA~\cite{CTIA} and TCPA~\cite{TCPA} guidelines regarding to unsolicited messaging.

Popular bulk texting applications, such as Twilio, serve as ESME which implement these considerations into their services by requiring consent before you are able to send out messages. CTIA best practices state that, "Message Senders should use reasonable efforts to prevent and combat unwanted or unlawful
messaging traffic, including spam and unlawful spoofing"~\cite{CTIA}, which includes the handling of smish. Although CTIA is not a legal document, TCPA is a set of legal rules that restrict sending unsolicited text messages. While legal efforts have outlawed smishing campaigns and these services follow such laws, smishing attacks are still sent out in bulk. During the process of selecting bulk messaging services for this research, we were notified by several companies that fraudulent users were a problem for their services. Clicksend~\cite{clicksend} mentioned in an email that, "Unfortunately, we get about 20 fraudulent users per day attempting to sign up to our service, so sometimes we have to flag new accounts for review" during the account creation process. Additionally, while creating an account, services may require a trial or approval process before they give access to their API. On one service, EZTexting~\cite{eztexting}, our initial messages were placed under review before we could send out bulk campaign messages. When API access was restricted, we applied for access and explained our research in the application; however, none of the selected bulk messengers required an approval process. Requiring users to build up trust before allowing them to access the API is one way to mitigate fraudulent messaging. Furthermore, we also noticed that while adoption of CTIA and TCPA was universal, additional limitations to block unwanted messages were implemented on most services. For example, Trumpia~\cite{trumpia} required an opt-out tag at the end of each message, or that campaign targets receive an opt-in for the first message of the campaign, before regular messaging could begin. In most circumstances, these restrictions could be removed by purchasing account upgrades. Additionally, several messaging services limit account access to only business clientele, which could make it very difficult for smishing attackers to bypass. Due to these circumstances, we selected bulk messaging services which were available for all users. 20 messaging services from top rated text marketing service lists~\cite{bulkservicelist, bulkmessengerlist} were explored. We chose our final 5 bulk messenger services based on price, access to an API, messaging restrictions, market share and user ratings. We believe that smishing campaigns will gravitate towards services with fewer messaging restrictions, allowing easier account creation and fewer blocked phishing messages from the services. Moving forward, we did not face difficulty in finding services which allowed non-business customers and could remove the opt-out messages. However, since each bulk messaging application uses different algorithms and heuristics, it is important to determine the steps taken by popular services to curb the sending of phishing messages. A list of the bulk messaging services considered in this research can be found in Table~\ref{tab:bulkserviceslist}.

To deliver the messages, we either accessed the developer API provided by each service or the built-in messaging services through their site. We then process our bulk smishing campaign through these services to send the 40 messages to each testing phone.
\vspace{-1mm}
\subsection{Overview of the Mobile Devices and Carriers}

\hspace{0.4cm} A list of carriers and their filtering systems considered in this work can be found in Table \ref{tab:carrier-table}. 
We note that the market share for MintMobile is not publicly available. While AT\&T, Verizon and T-Mobile consist of the majority share of subscription plans in the US, the inclusion of MetroPCS and MintMobile allows us to test some assumptions about the mobile carriers. 
Additionally, MetroPCS is owned by T-Mobile and its SMS protection application, while named slightly differently, is presented by T-Mobile as well. MintMobile on the other hand only shares the same network as T-Mobile and is owned separately. If there is a difference in blocking rates between MintMobile and T-Mobile, it will likely be based on the applications on the device as opposed to the carrier network blocking the message. Although MetroPCS is part of T-Mobile, they carry messages on the same network, and are owned by the same company, we treat them separately in this paper. This allows us to determine if there were any underlying services specific to T-Mobile and the systems that share its' network, as well as features that may differentiate their blocking capabilities. 
For this research, we only focus on the default protection provided by each carrier. Therefore, we ask our participants to set their protection to the default settings before sending messages to their device.

\textbf{Device factors considered:} 
We chose an equal share of Android and iOS devices, each corresponding to a mobile carrier with their default SMS protection settings. The filtering systems for each device are assigned randomly to avoid bias of one platform over the other. These filtering systems consist of built-in SMS security that comes pre-packed on the mobile devices by each carrier, and are matched by a corresponding anti-smishing app. 
As we send 40 messages to each device, we also consider the implications that the carrier may consider our activity on the devices as suspicious. Carrier violations when handling bulk texting may hinder our ability to obtain accurate results. Therefore, part of this research includes monitoring phone carriers, so that they do not flag our activity and prevent delivery of all messages, regardless of whether they are smishing or not. Similarly, we monitor the bulk messaging services, to not classify restrictions or bans on our account for a successful smish identification. To mitigate this possibility, we check the bulk messaging service in between each messaging campaign to ensure that services are still available.

\vspace{-1mm}
\subsection{Overview of the 3rd Party Apps}

\hspace{0.4cm} For this research, we selected 10 popular text filtering applications, split between Android and iOS. While few tools explicitly state they are anti-smishing apps, they do state that they filter SMS messages. 
The anti-smishing apps that we analyzed lack a specific filtering identification between spam and smishing in their text filtering. In other words, the apps bundle smishing and spam blocking under the same label and filter the messages without distinction. Work has been done to separate the distinction between spam and smishing in the security community, and many of the same features used to detect spam apply to smishing~\cite{smishingfromspam}. However, in this work we do not consider them the same. To avoid misidentification, our message set is limited to only smish or ham examples.
We test these apps using their basic SMS protection features. The features of each app are analyzed on their ability to filter messages. The approach to setting up these 3rd party apps primarily involves downloading them and then setting the 3rd party app as the default SMS app on the phone. For our analysis phase of the phishing message protection by the anti-smishing app, we isolate the effects of the carrier and the sender in the previous tests, and then determine the additional messages blocked. Additionally, in this paper we study the impact of false positives, as they negatively impact the end user and may contribute to their adoption of an anti-smishing tool. The 3rd party apps used in this study are listed in Table \ref{tab:3rd-party-apps}.

\subsection{Experiment Design and Criteria}

\hspace{0.4cm} To accurately isolate the effects of bulk senders, carriers, and the 3rd party apps, we will need to perform three experiments. 
Only the third experiment is conducted with third-party apps installed on the devices. We send our messages through each bulk messaging service to the same carrier. As bulk messaging services and carriers block SMS at different transmission ends, we must perform a second test using an experimental design to isolate the carrier impact on message blocking. When a carrier fails to receive a message, identification alone is not sufficient to determine why the message was blocked. However, by observing the effects of isolating the carrier and the sender, we can find differences in the block rates and attribute those to either one group or the other. If the carrier detects an illegitimate message, it returns a notification to the bulk texting service that the message was not delivered, along with notifying them that the message was blocked for containing malicious messages. For security purposes, these messaging services do not identify whether it was the carrier or the sender that filtered the message. When receiving a high volume of these responses from a carrier, it is likely that the bulk messaging service will limit or suspend our accounts. Therefore, we verify the submission status of messages through the service's provided logs or screenshots of the messages arriving at the intended target phone. After determining what messages a carrier and bulk messaging service will block, we apply the 3rd party apps and study how they filter the remaining messages. The third experiment will use a single carrier and anti-smishing app combination as the target for our messages. By doing so, we can track performance of each 3rd party app. While our experiment aims to mimic real-world smishing attacks, we acknowledge that we cannot thoroughly test all aspects of the anti-smishing technology of these systems. Smishing detection is not limited to the messages themselves but can include many factors such as the context of the sender and the messaging behavior. We attempt to minimize these effects by using fresh accounts from each messenger, thereby limiting the amount of sender history that could affect the sending rates. Similarly, we diversify the targets of our messaging to utilize unique phone targets for message delivery.
\section{Results}
\label{sec:results}

\hspace{0.4cm} We evaluate our study using the \textit{smish hit rate} and \textit{benign strike rate}. In our study, the higher the smish hit rate and the lower the benign strike rate, the better.

\begin{figure}[htbp]
    \vspace{-3mm}
    \centering
    \includegraphics[clip, trim=4cm 1cm 1cm 2cm, height=5.5cm]{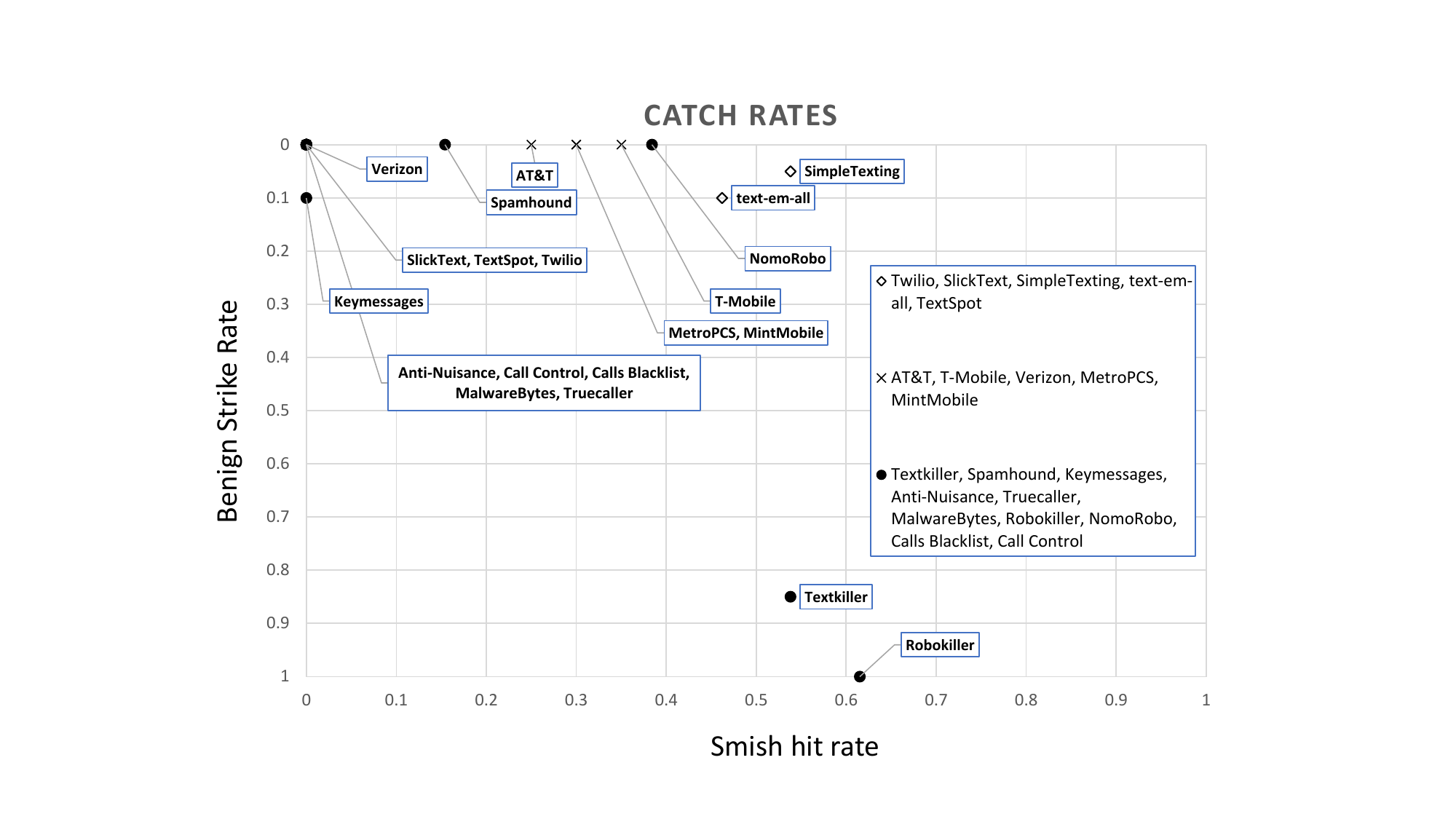}
    \caption{The smish hit rate and benign strike rate of Messengers, Carriers and Apps. The right-hand upper corner of the figure indicates the best performance.}
    \label{fig:Allhitandstrikerate}
    \vspace{-3mm}
\end{figure}

\subsection{Smish Hit Rate of Bulk Messaging Services}
\label{subsec:bulkservice}

\hspace{0.4cm} In this testing step, we separate the blocking rates of our bulk messaging services from our carriers. A graph of the smish hit rate and benign strike rate for messages sent by each service is illustrated in Figure~\ref{fig:Allhitandstrikerate}.
We found that some bulk messaging services performed significantly different from others at handling our messages. Of the services tested, SimpleTexting delivered the fewest phishing messages with a smish hit rate of 53.8\%.  
60\% of the tools performed the same against our messaging set. We checked the logs on these services and found that all messages have been marked as sent. Based on this, we identify that the messages were blocked by the carrier and not the bulk messaging service. Similarly, 35\% of the smish hit rate blocked the same messages across all bulk messaging services, which we attribute to the T-Mobile carrier. We found that 25\% of the phishing messages were able to be delivered by all services to target phones. The results of a chi-square of the test show that SimpleTexting (${\chi}^2 (1, N = 26)$ = 9.579, \textit{p} = .002) and text-em-all (${\chi}^2 (1, N = 26)$ = 7.800, \textit{p} = .005) performed significantly better at catching phishing messages than Twilio, SlickText and TextSpot. However, we did not find a significant difference between SimpleTexting and text-em-all.

When we considered correctly identifying all messages, we did not find any significant difference between SimpleTexting or text-em-all and other messaging services.
 On the other hand, the highest benign strike rate of the bulk messaging services was 10\% for text-em-all. Fisher's exact test was used to compare SimpleTexting to Twilio,TextSpot or SlickText and found that there was no statistically significant difference between them(one-tailed \textit{p} = .244). While a couple benign messages were blocked by two of the senders, we do not anticipate that this is a major issue among bulk messaging services. The purpose of these services is typically for marketing messages instead of allowing conversational benign SMS dialogue messages through their service. This shows room for improvement among bulk messaging services in blocking SMS phishing messages. It is important to note that several messaging service responded quickly to our smishing campaigns. Twilio, Slicktext and SimpleTexting suspended our accounts within 48 hours of starting the smishing campaigns. We were notified of the suspensions through our email addresses tied to the accounts. However, we had sent all of our messages at the point whern our account had been banned. As previously stated, bulk messaging services can be notified when a carrier blocks an SMS. Even if a bulk messaging service missed our smishing attacks themselves, a carrier could flag our smishing campaign and relay that information back to the bulk messaging service to close our account. 
 Additionally, once we identified which smish had been blocked by which carrier, we could have tailored a campaign to send smishing messages that would not have been flagged by either carrier network or bulk messaging service.

\subsection{Smish Hit Rate of Mobile Carriers}
\label{subsec:mobilecarrier}

\hspace{0.4cm} Evaluation of the smish hit rate and benign strike rate of carriers was done by isolating them as a testing variable and relying on the default message blocking applications that are installed on each phone. We selected Twilio as our bulk messaging service and sent the same selection of 40 messages to each carrier. The goal of this experimental design was to separate the effects the bulk messenger service and carrier had on messages blocked at the receiving phone. The blocking rate of our carriers can be viewed in Figure~\ref{fig:Allhitandstrikerate}.
The list of messages referenced here is also available in the Appendix. 
Based on our results, we identified that MintMobile, MetroPCS, and T-Mobile, which share the T-Mobile network, did not block the same messages. While similar messages are blocked, we noticed that the T-Mobile service was blocking one additional phishing message.
This was verified by an additional test on the carrier service. 
Moreover, this additional blocked message may be attributed to carrier specific services.
The highest smish hit rate of carriers tested is 35\% by T-Mobile. 
Alternatively, Verizon network delivered all messages sent from Twilio during our testing, including both smishing and benign messages, to the target device. By verifying all messages were delivered to Verizon, we can confirm that none of our messages are being detected as spam by the Twilio service itself, and that any message blocked in this stage is due to the carrier. As expected, the benign strike rate of carriers is 0\%. Using Fisher's exact test to compare any two carriers on the basis of benign strike rate shows that no carrier performs significantly different in striking benign messages(one-tailed p = 1.00); however, the large number of phishing messages that can be sent to our carriers is an interesting finding. Based on a chi-square test of phishing messages we determine that T-Mobile (${\chi}^2 (1, N = 40)$ = 8.485, \textit{p} = .004), MetroPCS, MintMobile (${\chi}^2 (1, N = 40)$ = 7.059, \textit{p} = .008) and AT\&T (${\chi}^2 (1, N = 40)$ = 5.714, \textit{p} = .017) performed significantly better than Verizon. However, we did not find a significant difference between T-Mobile and AT\&T, MintMobile or MetroPCS. While T-Mobile performed best in this experimental stage, it still allows 65\% of our smsishing messages to be delivered to their carrier devices. Additionally, when we considered both benign and phishing messages, T-Mobile 
did not perform significantly different than Verizon. Based on our results, we can confidently determine which messages are being blocked by third party apps.

\subsection{Smish Hit Rate of Anti-Smishing Apps}
\label{subsec:antiapp}

\hspace{0.4cm} The blocking rate of each anti-smishing app is determined by sending messages to our apps through the T-Mobile network. By sending through T-Mobile we exclude messages from consideration that are already blocked by the carrier, as we are comparing the differences that apps make on detection. 
Although T-Mobile and MetroPCS did not block the same messages, we are confident that including MetroPCS in this phase as a target carrier will not bias the messages. This is because the messages removed from consideration include all the messages blocked by MetroPCS. Thus all messages analyzed in this section will be messages that are not blocked by MetroPCS. 
All smishing apps are tested using the message blocking option with their default settings. Figure~\ref{fig:Allhitandstrikerate} shows the block rate for the apps, both in terms of the benign strike rate and smish hit rate. While the messages identified in the previous carrier test were not considered in these results, we note that the same messages were blocked by all apps, showing their blocking was consistent.
Regarding additional catch rates of phishing messages, we found that Textkiller, Robotkiller and NomoRobo had the highest blocking rates. Of the three, we disocvered that Robokiller blocked the most phishing messages with a smish strike rate of 61.6\% for the message set. However, it still leaves 38.4\% of phishing messages sent to the target phone. Additionally, 60\% of anti-smishing apps had a smishing hit rate of 0 on top of messages that were blocked by the carrier service. The next best service, TextKiller had a smish hit rate of 53.9\% for the message set. Through a chi-square test we found that Robokiller (${\chi}^2 (1, N = 26)$ = 11.556, \textit{p} $<$ .001) and TextKiller (${\chi}^2 (1, N = 26)$ = 9.579, \textit{p} = .002) perform significantly better than Call Control, Calls Blacklist, MalwareBytes, TrueCaller, and Anti-Nuisance at blocking phishing messages. Additionally, Robokiller (${\chi}^2 (1, N = 26)$ = 5.850, \textit{p} = .016) and TextKiller (${\chi}^2 (1, N = 26)$ = 4.248, \textit{p} = .039) performed significantly better than Spamhound against our phishing messages. However, we did not find a statistically significant difference between Textkiller, Robokiller and NomoRobo in blocking phishing messages.

When considering the legitimate messages along with the phishing messages, we found that the anti-smishing tools, which were able to filter a large portion of the remaining phishing messages, also filtered some legitimate messages. In particular, Robokiller, which performed the best at filtering phishing messages, had a benign strike rate of 100\%. Interestingly, for Robokiller, we had a higher rate of phishing messages delivered to the target device than legitimate ones. A chi-square test revealed that both TextKiller (${\chi}^2 (1, N = 66)$ = 13.687, \textit{p} $<$ .001) and Robokiller (${\chi}^2 (1, N = 66)$ = 17.515, \textit{p} $<$ .001) performed significantly worse than NomoRobo against the entire message set. Separately, we found that Robokiller blocked many messages when the sender number was unknown. However, when the sender number was added as a contact, the messages shifted from the spam filter to the inbox. This implies a catch-all approach to message detection instead of independently checking each message for smishing characteristics. When comparing smishing and benign, the best performing anti-smishing app was NomoRobo which accurately filtered 75.8\% of our messages.  
\subsection{Responsibility Disclosure.} We reached out to all Bulk Messaging Services, Carriers and Anti-Smishing Apps used in this experiment to share our research findings. Currently, we have received responses from Textspot, Twilio, NomorRobo and Truecaller who thanked us for contacting them with this information. We further spoke to the Nomorobo CEO who explained, \textit{"It’s great to get more eyes on fighting SMS fraud. As the researchers have shown, and from the trenches, we know, this is a really hard problem to solve."}.
\section{Discussion}
\label{sec:discussion}

\paragraph{Discussion of Messages Blocked}
Previously, we identified the catch rate of smishing smishing attacks by our tools. Among bulk messaging services, we found that most did not block any additional messages over the target carrier services. However, from those bulk messaging services that did block additional messages, 62.5\% of the new blocked messages were blocked on SimpleTexting and Text-em-all. In our testing, we identified 1 spear phishing SMS; however, it was also the most consistently blocked message across all bulk messaging services, carriers and anti-smishing apps.

None of the messages were found to be commonly blocked across all carriers. However, 30\% of our phishing messages were blocked by all the carriers that used the same T-Mobile network. Additionally, we found that amongst our carriers there were some common factors in the messages being blocked across tools. For instance, two of the most commonly blocked messages are variations on a very common phishing message scam~\cite{fifththirdscam,freemessagescam}. Lastly, we noticed some anti-smishing apps were incorrectly filtering many of our benign messages. Particularly, Robokiller blocked all of our benign messages while leaving 38.4\% of the phishing messages in the target's inbox. Only NomoRobo was able to make a large impact of an additional 38.5\% of phishing messages blocked without simultaneously blocking legitimate texts. A further investigation into the phishing messages that were allowed through Robokiller shows that they all contain links to websites that were no longer active during the testing phase. This may provide insight into why they were not filtered. Across the apps that specifically report using a blacklist in their smishing detection~\cite{androidstoreCBL,iosTK}, we did not find commonality among the messages blocked.

\paragraph{Discussion of Tool performances}
 We found that the tested tools performed poorly against our smishing attacks. This highlights the need for improvement among anti-smishing tools, especially against newer attacks. We identified commonalities amongst messages blocked by carriers; however, apps and messaging services did not show significant overlap in the messages they blocked. The commonality among messages blocked by MintMobile, MetroPCS and T-Mobile implies a similar filtering performance over shared carrier networks. Separately, the differences in messages blocked between apps and messaging services show that they are likely using different techniques to block messages. An aspect of the bulk messaging services is their compliance with CTIA~\cite{CTIA} and TCPA~\cite{TCPA}. We found that this compliance did not effectively deter us from sending unsolicited phishing messages through the bulk messaging services. In our testing, we were able to send our texts to each target phone without initially providing any proof of consent. In all tested bulk messaging services, besides text-em-all, we were also able to remove the opt-out message from all texts through purchasing subscriptions or top-up payments. Even among services that require opt-out, phishing messages exist which incorporate this opt-out message in their scam. Comparatively, the carrier that performed the best against our phishing messages was T-Mobile, while Verizon failed to catch any of our smishing attacks. 
 The resulting catch rates for carrier services show that our smishing set has a high delivery rate against modern anti-smishing protections. We see room for improvement among these services in their ability to catch new smishing attacks while also not filtering benign messages. Finally, among the anti-smishing apps, we found that most of the tools provided no additional message blocking over the built-in carrier blocking. Of the tools that provided additional protection, we found many benign messages being blocked. 

\vspace{-2mm}
\subsection{Recommendations}

\hspace{0.4cm} Based on our testing of bulk messaging services, we found that we were able to remove or avoid the opt-out text from our messages from all services, besides text-em-all. Many of these services require this to be included initially, but with a purchase of credits or a subscription these could be removed. Additionally, before sending our campaigns, we did not need to gather or show proof of any opt-in permissions for all services. 
All bulk messaging services in this study state that they follow the CTIA~\cite{CTIA}, in sections 5.1.1 and 5.1.2 of their guidelines they describe policies for opt-in and opt-out communications. Based on current US legal precedence, there is ambiguity over whether bulk messaging services, such as Twilio, are liable for the TCPA violations of its users~\cite{TCPATwilio}. We recommend bulk messaging services require their customers to receive an opt-in response before campaigns and a clear opt-out with future messaging campaigns. 

To deliver our smishing campaign, we first explored the option to send messages through an API. 
We found that access to this API was rarely restricted and, in most cases, we could sign up and begin using it immediately. Alternatively, we found messaging services that restricted their API typically use an approval process. This approval process came in many forms, but we found the most common one to be a verification that we represent a legitimate business. To send the smishing campaign on a large scale with the method that we used in this study, it would require API access, and this could be a significant hurdle for an attacker. We recommend for this to become a requirement among all bulk messaging services. This can be done by verifying that a user is operating an authentic website, and that they have a valid business before granting access to web services.

Apps should clarify to the sender why messages are being blocked, and filtering should remove messages from the inbox. When testing anti-smishing apps, we noticed that some apps classified our messages as sales, or unverified, while still leaving them in the inbox with other messages. These seemed to be catch-all terms for suspicious messages. Our methodology does not count those as successfully filtered phishing messages. We recommend that app filtering make clear distinctions between all types of messages and separate them from regular messages.

\subsection{Limitations}

\hspace{0.4cm} While this work provides insight into the blocking and detection rates of popular anti-smishing tools, it has several limitations. 
While we were able to test several tools in this study, we recognize that a full permutation of every message to every combination of bulk texting service, mobile carrier app, and the anti-smishing app would be infeasible for the scope of this study. Instead, we opted to perform three separate tests to identify the contributions of each tool. This method identifies the generalized blocking rate of each tool; however, it may fail to identify edge cases. Additionally, while the specific techniques used by each tool would be interesting to explore, this information is not publicly available and is outside the premise of this work. 

Our research aimed to use fresh smishing messages. While we were able to collect recent smish, we were limited in how quickly we could deliver them. Due to the potential for disruptions in the delivery as a result of message reviews, we opted to send our data set collectively instead of as received. This led to our smishing set containing messages of varying degrees of freshness. As delivery of the messages took 3 days, there is a potential for filtering or network paths to change. However, with the consistency of messages blocked between tests, we believe this did not affect our results.

Our study design uses phishing messages collected from the public through the smishtank website. This method requires a user to first recognize that a message may be a smishing attack, and then post it onto our website for validation. While this approach is robust in its ability to collect a large number of phishing messages, some types of messages are more difficult to detect than others. We recognize that this may imbalance our message collection towards phishing messages which are easier to spot.

\subsection{Future work}

\hspace{0.4cm}Through our study, we recognized the lack of recognition of ML smishing detection approaches. 
While related works have discovered many machine learning approaches to tackle zero-day smishing attacks, few references were made to ML in modern commercial anti-smishing tools. Zero-day smishing attacks continue to remain the largest struggle for anti-smishing tools. Consequently, future work should be done to investigate the involvement of machine learning not just in the research community, but in the general commercial products.

We studied 3 different variables in the smish blocking process. 
However, there are more places that could serve to block messages. 
All SMS messages pass through SMSC, and analyzing the techniques employed at the service center level could provide important insights into smishing prevention. This study revealed that carriers on the same network blocked similar messages. This could be further broken down between the services on carrier networks. 

In this study, we focused solely on SMS messages. 
However, there are other forms of text messaging services that employ anti-smishing technology. 
Work on messaging services like Over The Top, which have similar properties or SMS, could provide important information on mobile communications. Additionally, we wish to explore the difference in delivery rates when sending from local, toll-free numbers, and email to text.
\section{Conclusion}
\label{sec:conclusion}

\hspace{0.4cm} Smishing attacks are on the rise, and anti-smishing tools have been created to combat these threats. We attempted to provide a source for public smishing data sets through our contribution of smishtank.com. We then analyzed our collected messages through VirusTotal and characterized them based on squatting techniques, URL types, named entities and subcategories. In this paper, we compared the effectiveness of anti-smishing technology of 5 bulk messaging services, 5 carriers and 10 anti-smishing apps across three separate tests. There are a variety of techniques provided by popular anti-smishing tools, some disclosed by the services while others remain hidden. We not only investigated whether messages were being blocked, but also at what point in the chain of communication that messages were being detected and filtered. Our comparative analysis of these tools found room for improvement against new smishing attacks, and our results provide an overview of areas of improvement. 
\section{Acknowledgement}

The authors are grateful to all the participants who donated their time to submit smishing samples to smishtank.com and participated in our research. We also thank Soha Khoso for proofreading the draft version of the paper. We appreciate WiSec'23 anonymous reviewers and Shepherd for their constructive feedback and comments.

\bibliographystyle{plain}
\bibliography{main}

\begin{thebibliography}{10}

\bibitem{impactofantiphishingtools}
Ahmed Abbasi, Fatemeh Zahedi, and Yan Chen.
\newblock Impact of anti-phishing tool performance on attack success rates.
\newblock In {\em 2012 IEEE International Conference on Intelligence and
  Security Informatics}, pages 12--17, 2012.

\bibitem{smsspamfilteringcollection}
Tiago~A. Almeida, Jos\'{e} Mar\'{\i}a~G. Hidalgo, and Akebo Yamakami.
\newblock Contributions to the study of sms spam filtering: New collection and
  results.
\newblock In {\em Proceedings of the 11th ACM Symposium on Document
  Engineering}, DocEng '11, page 259–262, New York, NY, USA, 2011.
  Association for Computing Machinery.

\bibitem{androidstoreCBL}
{Calls Blacklist} - call blocker.
\newblock
  \url{https://play.google.com/store/apps/details?id=com.vladlee.easyblacklist}.

\bibitem{apwg}
Apwg | unifying the global response to cybercrime.
\newblock https://apwg.org/.

\bibitem{avoidphishbait}
Said Baadel, Fadi Thabtah, and Asim Majeed.
\newblock Avoiding the phishing bait: The need for conventional countermeasures
  for mobile users.
\newblock In {\em 2018 IEEE 9th Annual Information Technology, Electronics and
  Mobile Communication Conference (IEMCON)}, pages 421--425, 2018.

\bibitem{bitly}
{URL Shortener - Short URLs \& Custom Free Link Shortener}.
\newblock \url{https://bitly.com/}.

\bibitem{fifththirdscam}
Sarah Brookbank.
\newblock Don't click that link: Scammers are targeting fifth third bank
  customers.
\newblock
  \url{https://www.cincinnati.com/story/news/2020/12/23/dont-click-link-scammers-targeting-fifth-third-bank-customers/4005789001/}.

\bibitem{freemessagescam}
Antonio Camacho.
\newblock A spam text from your own number? don't get phished - cnet.
\newblock
  \url{https://www.cnet.com/tech/mobile/a-spam-text-from-your-own-number-dont-get-phished/}.

\bibitem{honeypot}
Shubhika Chauhan and Savita Shiwani.
\newblock A honeypots based anti-phishing framework.
\newblock In {\em 2014 International Conference on Control, Instrumentation,
  Communication and Computational Technologies (ICCICCT)}, pages 618--625,
  2014.

\bibitem{surveyantiphishtechniques}
Sharvari~Prakash Chorghe and Narendra Shekokar.
\newblock A survey on anti-phishing techniques in mobile phones.
\newblock In {\em 2016 International Conference on Inventive Computation
  Technologies (ICICT)}, volume~2, pages 1--5, 2016.

\bibitem{cisco}
Cisco.
\newblock Cyber security threat trends: phishing, crypto top the list, 2021
  (accessed April 10, 2022).
\newblock
  https://learn-umbrella.cisco.com/ebook-library/2021-cyber-security-threat-trends-phishing-crypto-top-the-list.

\bibitem{clicksend}
Clicksend | business sms, voice, mail and more via web or api.
\newblock \url{https://www.clicksend.com/us/}.

\bibitem{CTIA}
CTIA.
\newblock Messaging principles and best practices, 2021 (accessed April 17,
  2022).
\newblock
  \url{https://api.ctia.org/wp-content/uploads/2019/07/190719-CTIA-Messaging-Principles-and-Best-Practices-FINAL.pdf}.

\bibitem{phishmonger}
David~G. Dobolyi and Ahmed Abbasi.
\newblock Phishmonger: A free and open source public archive of real-world
  phishing websites.
\newblock In {\em 2016 IEEE Conference on Intelligence and Security Informatics
  (ISI)}, pages 31--36, 2016.

\bibitem{eztexting}
{Text Marketing} - sms marketing – 2020 best mms marketing software.
\newblock \url{https://www.eztexting.com/}.

\bibitem{localnumberbetter}
EZTexting.
\newblock {What is an A2P-enabled high-volume number and why should I upgrade?}
\newblock
  \url{https://eztexting.force.com/answers/s/article/What-is-an-A2P-number-and-why-should-I-upgrade},
  2022.
\newblock [Online; accessed 22-December-2022].

\bibitem{fbi}
FBI.
\newblock Federal bureau of investigation intenert crime report 2021, 2021
  (accessed April 10, 2022).
\newblock \url{https://www.ic3.gov/Media/PDF/AnnualReport/2021\_IC3Report.pdf}.

\bibitem{TCPA}
FCC.
\newblock Telephone consumer protection act 47 u.s.c. § 227, 2021 (accessed
  April 17, 2022).
\newblock \url{https://www.fcc.gov/sites/default/files/tcpa-rules.pdf}.

\bibitem{bulkservicelist}
Max Freedman.
\newblock The best text message marketing services of 2022.
\newblock
  \url{https://www.businessnewsdaily.com/15044-best-text-message-marketing-solutions.html}.

\bibitem{bulkmessengerlist}
G2.
\newblock Best sms marketing software in 2022: Compare reviews on 290+ | g2.
\newblock \url{https://www.g2.com/categories/sms-marketing}.

\bibitem{vitalworks}
Natalie Goguen.
\newblock Vitalworks and microsoft reach settlement.
\newblock
  \url{https://www.noip.com/blog/2014/07/09/vitalwerks-microsoft-reach-settlement/},
  2014.
\newblock [Online; accessed 1/6/23].

\bibitem{smsgatewayfirewall}
Abhijeet Guha.
\newblock Sms firewall – the feature you need in an smsc.
\newblock \url{https://www.revesoft.com/blog/sms-platform/sms-firewall/}.

\bibitem{IDACHABA201496}
Francis~Enejo Idachaba.
\newblock Algorithm for source mobile identification and deactivation in sms
  triggered improvised explosive devices.
\newblock {\em Procedia Engineering}, 78:96--101, 2014.
\newblock Humanitarian Technology: Science, Systems and Global Impact 2014,
  HumTech2014.

\bibitem{iosTK}
{TextKiller} spam text blocker.
\newblock
  \url{https://apps.apple.com/us/app/textkiller-spam-text-blocker/id1514005355}.

\bibitem{smishincrease}
itpro.
\newblock {Smishing attacks increased 700\% in first six months of 2021}.
\newblock
  \url{https://www.itpro.com/security/scams/360873/smishing-attacks-increase-700-percent-2021},
  2021.
\newblock [Online; accessed 24-July-2022].

\bibitem{JAIN2018617}
Ankit~Kumar Jain and B.B. Gupta.
\newblock Rule-based framework for detection of smishing messages in mobile
  environment.
\newblock {\em Procedia Computer Science}, 125:617--623, 2018.
\newblock The 6th International Conference on Smart Computing and
  Communications.

\bibitem{krebs_2021}
Brian Krebs.
\newblock U.k. arrest in 'sms bandits' phishing service, Feb 2021.

\bibitem{phishbox}
Jhen-Hao Li and Sheng-De Wang.
\newblock Phishbox: An approach for phishing validation and detection.
\newblock In {\em 2017 IEEE 15th Intl Conf on Dependable, Autonomic and Secure
  Computing}, pages 557--564, 2017.

\bibitem{smsphishingandmitigation}
Sandhya Mishra and Devpriya Soni.
\newblock Sms phishing and mitigation approaches.
\newblock In {\em 2019 Twelfth International Conference on Contemporary
  Computing (IC3)}, pages 1--5, 2019.

\bibitem{Mishra2021DSmishSMSAST}
Sandhya Mishra and Devpriya Soni.
\newblock Dsmishsms-a system to detect smishing sms.
\newblock {\em Neural Computing \& Applications}, pages 1 -- 18, 2021.

\bibitem{smishingfromspam}
Cik~Feresa Mohd~Foozy, Rabiah Ahmad, and Mohd Abdollah.
\newblock A practical rule based technique by splitting sms phishing from sms
  spam for better accuracy in mobile device.
\newblock {\em International Review on Computers and Software (IRECOS)},
  9:1776, 10 2014.

\bibitem{blacklisteffectiveness}
Adam Oest, Yeganeh Safaei, Adam Doupé, Gail-Joon Ahn, Brad Wardman, and Kevin
  Tyers.
\newblock Phishfarm: A scalable framework for measuring the effectiveness of
  evasion techniques against browser phishing blacklists.
\newblock In {\em 2019 IEEE Symposium on Security and Privacy (SP)}, pages
  1344--1361, 2019.

\bibitem{insidephishersmind}
Adam Oest, Yeganeh Safei, Adam Doupé, Gail-Joon Ahn, Brad Wardman, and Gary
  Warner.
\newblock Inside a phisher's mind: Understanding the anti-phishing ecosystem
  through phishing kit analysis.
\newblock In {\em 2018 APWG Symposium on Electronic Crime Research (eCrime)},
  pages 1--12, 2018.

\bibitem{onlinethreatalerts}
Online Threat~Alerts (OTA).
\newblock {Costco Text Scam Reward Coupon and Airpod Raffle}.
\newblock
  \url{https://www.onlinethreatalerts.com/article/2021/6/13/costco-text-scam-reward-coupon-and-raffle/},
  2022.
\newblock [Online; accessed 12/25/22].

\bibitem{phishtank}
{Join the fight against phishing} phishtank.
\newblock \url{https://phishtank.org/}.

\bibitem{phishnet}
Pawan Prakash, Manish Kumar, Ramana~Rao Kompella, and Minaxi Gupta.
\newblock Phishnet: Predictive blacklisting to detect phishing attacks.
\newblock In {\em 2010 Proceedings IEEE INFOCOM}, pages 1--5, 2010.

\bibitem{proofpoint}
proofpoint.
\newblock 2022 state of the phish, 2022 (accessed April 10, 2022).
\newblock
  \url{https://www.proofpoint.com/us/resources/threat-reports/state-of-phish}.

\bibitem{bethephisher}
Florian Quinkert, Martin Degeling, Jim Blythe, and Thorsten Holz.
\newblock Be the phisher -- understanding users' perception of malicious
  domains.
\newblock In {\em Proceedings of the 15th ACM Asia Conference on Computer and
  Communications Security}, ASIA CCS '20, page 263–276, New York, NY, USA,
  2020. Association for Computing Machinery.

\bibitem{smsbulkmessaging}
Bradley Reaves, Logan Blue, Dave Tian, Patrick Traynor, and Kevin~R.B. Butler.
\newblock Detecting sms spam in the age of legitimate bulk messaging.
\newblock In {\em Proceedings of the 9th ACM Conference on WiSec}, WiSec '16,
  page 165–170, New York, NY, USA, 2016. ACM.

\bibitem{sendingoutsms}
Bradley Reaves, Nolen Scaife, Dave Tian, Logan Blue, Patrick Traynor, and Kevin
  R.~B. Butler.
\newblock Sending out an sms: Characterizing the security of the sms ecosystem
  with public gateways.
\newblock In {\em 2016 IEEE Symposium on Security and Privacy (SP)}, pages
  339--356, 2016.

\bibitem{smspublicgateways}
Bradley Reaves, Luis Vargas, Nolen Scaife, Dave Tian, Logan Blue, Patrick
  Traynor, and Kevin R.~B. Butler.
\newblock Characterizing the security of the sms ecosystem with public
  gateways.
\newblock {\em ACM Trans. Priv. Secur.}, 22(1), dec 2018.

\bibitem{comparativeanalysisphishtools}
Himani Sharma, Er. Meenakshi, and Sandeep~Kaur Bhatia.
\newblock A comparative analysis and awareness survey of phishing detection
  tools.
\newblock In {\em 2017 2nd IEEE International Conference on Recent Trends in
  Electronics, Information Communication Technology (RTEICT)}, pages
  1437--1442, 2017.

\bibitem{smsclasstopicmodeling}
Dilip~Singh Sisodia, Shreya Mahapatra, and Arpita Sharma.
\newblock Automated sms classification and spam analysis using topic modeling.
\newblock In {\em 2nd International Conference on Data, Engineering and
  Applications (IDEA)}, pages 1--6, 2020.

\bibitem{cluesintwitter}
Siyuan Tang, Xianghang Mi, Ying Li, XiaoFeng Wang, and Kai Chen.
\newblock Clues in tweets: Twitter-guided discovery and analysis of sms spam,
  2022.

\bibitem{TCPATwilio}
{Twilio Texting Platform TCPA Class Action}.
\newblock
  https://www.natlawreview.com/article/platform-provider-paradox-text-platform-provider-twilio-may-be-directly-liable-tcpa.

\bibitem{trumpia}
{\#1 Texting Software} | sms automation software.
\newblock \url{https://www.trumpia.com/}.

\bibitem{UCICollection}
{SMS Spam Collection Data Set from UCI Machine Learning Repository} uci
  archive.
\newblock \url{http://archive.ics.uci.edu/ml/datasets/SMS+Spam+Collection}.

\bibitem{virustotal}
{VirusTotal} - home.
\newblock \url{https://www.virustotal.com/gui/home/url}.

\bibitem{taxonomy}
Rania Zaimi, Mohamed Hafidi, and Mahnane Lamia.
\newblock Survey paper: Taxonomy of website anti-phishing solutions.
\newblock In {\em 2020 Seventh International Conference on Social Networks
  Analysis, Management and Security (SNAMS)}, pages 1--8, 2020.

\bibitem{surveyofphishtools}
Hiba~Zuhair Zeydan, Ali Selamat, and Mazleena Salleh.
\newblock Survey of anti-phishing tools with detection capabilities.
\newblock In {\em 2014 International Symposium on Biometrics and Security
  Technologies (ISBAST)}, pages 214--219, 2014.

\bibitem{zhang2007phinding}
Yue Zhang, Serge Egelman, Lorrie Cranor, and Jason Hong.
\newblock Phinding phish: Evaluating anti-phishing tools.
\newblock {\em Proc. 14th Annual Network andDistributed System Security
  Sympos.1–16}, 2007.

\end{thebibliography}

\clearpage
\onecolumn
\section*{Appendix}

\appendix

\section{Bulk Messaging Services, Mobile Carriers \& Apps}

\begin{table}[ht]
\tiny
\vspace{-2mm}
\parbox{.50\linewidth}{
    \centering
    \notsotiny
    \begin{tabular}{@{}ccccc@{}}
    \toprule
    \multicolumn{5}{c}{Bulk Messaging Services}                              \\ \midrule
    Twilio & SlickText & SimpleTexting & text-em-all & TextSpot \\ \bottomrule
    \end{tabular}
    \caption{Bulk messaging services.}
    \label{tab:bulkserviceslist}
    \notsotiny
    \hfill
    \begin{tabular}{@{}lllll@{}}
    \toprule
    \multicolumn{5}{c}{Carrier{[}\# of Subscribers{]}} \\ \midrule
    T-Mobile{[}108.7M{]\tablefootnote{https://s29.q4cdn.com/310188824/files/doc\_financials/2021/q4/TMUS-12\_31\_2021-EX-99.1-vFinal.pdf}} & Verizon{[}142.8M{]\tablefootnote{https://www.verizon.com/about/file/60483/download?token=aC4JqAk6}} & AT\&T{[}\textgreater{}100M{]\tablefootnote{https://about.att.com/story/2022/q4-2021-results.html}} & Mint Mobile{[}N/A{]} & MetroPCS{[}N/A*{]} \\ \bottomrule
    \end{tabular}
    \caption{Mobile Carriers used in this study. * MetroPCS included in T-Mobile subscriber numbers.}
    \label{tab:carrier-table}
}\hfil
\parbox{.44\linewidth}{
    \centering
    \notsotiny
    \begin{tabular}{|ll|ll|}
    \hline
    \multicolumn{2}{|c|}{\textbf{Android}}                   & \multicolumn{2}{c|}{\textbf{iOS}}                       \\ \hline
    \multicolumn{1}{|c|}{\textbf{Name}} & \textbf{\#Ratings(Downloads)} & \multicolumn{1}{c|}{\textbf{Name}} & \textbf{\#Ratings(Downloads)} \\ \hline
    Key Messages    & 73.5K(1M+)\tablefootnote{https://play.google.com/store/apps/details?id=com.smsBlocker}   & SpamHound         & 0.7k (*)\tablefootnote{https://apps.apple.com/app/id1263185195}  \\
    Anti Nuisance   & 79.2K(1M+)\tablefootnote{https://play.google.com/store/apps/details?id=org.whiteglow.antinuisance}    & Robokiller   & 350.1K (*)\tablefootnote{https://apps.apple.com/us/app/robokiller-block-spam-calls/id1022831885} \\
    Call Control    & 109.4K(5M+)\tablefootnote{https://play.google.com/store/apps/details?id=com.flexaspect.android.everycallcontrol}  & Malwarebytes & 30.8K (*)\tablefootnote{https://apps.apple.com/US/app/id1327105431?mt=8}  \\
    Trucaller       & 17.35M(500M+)\tablefootnote{https://play.google.com/store/apps/details?id=com.truecaller} & Textkiller   & 9.7K (*)\tablefootnote{https://apps.apple.com/us/app/textkiller-spam-text-blocker/id1514005355}   \\
    Calls Blacklist & 747.5K(10M+)\tablefootnote{https://play.google.com/store/apps/details?id=com.vladlee.easyblacklist}  & NomoRobo     & 15.0K (*)\tablefootnote{https://apps.apple.com/us/app/nomorobo-robocall-blocking/id1134727588} \\ \hline
    \end{tabular}
    \caption{3rd Party Apps used in this study. * downloads data is not publicly available.}
    \label{tab:3rd-party-apps}
}
\end{table}

\vspace{-5mm}
\section{Message procedure \& Research variables}

\begin{figure}[htbp]
\vspace{-2mm}
    \begin{minipage}{.45\textwidth}
    \centering
        \includegraphics[width=.98\textwidth,height=4cm]{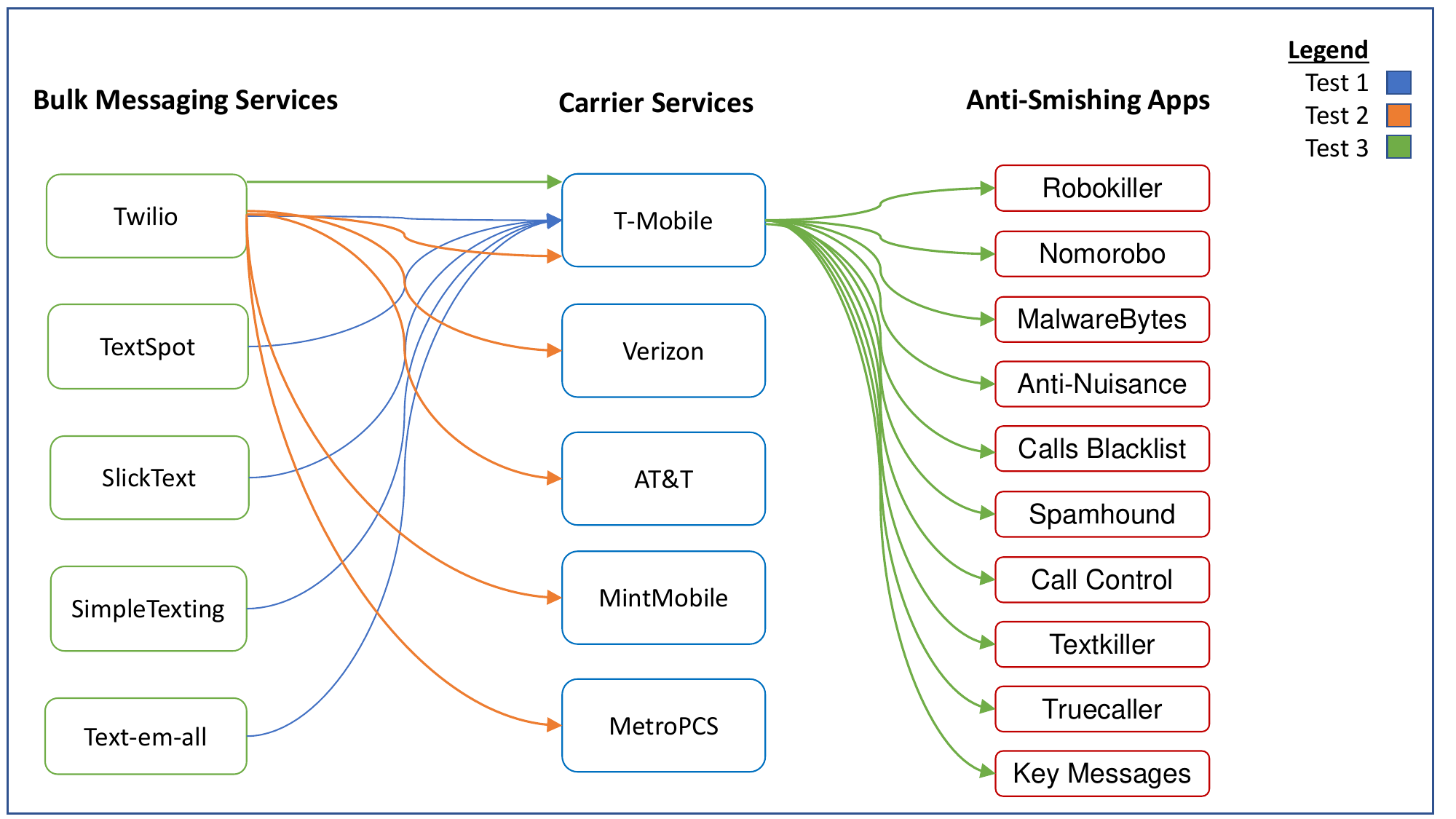}
    \caption{A visualization of messaging procedure.}
    \label{fig:testingmethodology}
    \vspace{-5mm}
    \end{minipage}
    \begin{minipage}{.54\textwidth}
    \vspace{-5mm}
    \centering
    \notsotiny
    \begin{tabular}{lll}
    \toprule
    No. & Variable & Description \\ \midrule
     & Independent Variables &  \\  \midrule
    1 & Bulk Messaging Services & The services we use to deliver our messages \\
    2 & Carrier & Telecommunication service providers that facilitate SMS messaging \\
    3 & Anti-smishing Apps & Apps which provide anti-smishing message filtering services \\  \midrule
     
     & Dependent Variables &  \\ \midrule
    4 & Smish Hit Rate & The true positive rate(TPR) of smishing messages correctly identified \\
    5 & Benign Strike Rate & The false positive rate(FPR) of benign messages incorrectly identified \\  \midrule
    \end{tabular}
    \caption{The variables and their descriptions used in this research.}
    \label{tab:variable-table}
    \vspace{-5mm}
    \end{minipage}
\end{figure}

\section{Data Characterization}

\begin{table}[ht]
\vspace{-3mm}
\centering
\setlength{\tabcolsep}{1.0pt}
\resizebox{\textwidth}{!}{%
\begin{tabular}{@{}lcccccccccccccccccccccc@{}}
\toprule
\multicolumn{3}{|l|}{} & \multicolumn{5}{c|}{Carrier Networks} & \multicolumn{5}{c|}{Bulk Messaging Services} & \multicolumn{10}{c|}{Anti-Smishing Apps} \\ \midrule
\multicolumn{1}{|c}{\begin{tabular}[c]{@{}c@{}}Data \\ Characterization\end{tabular}} & \begin{tabular}[c]{@{}c@{}}Messages\\ (All)\end{tabular} & \multicolumn{1}{c|}{\begin{tabular}[c]{@{}c@{}}Messages\\ (Selected)\end{tabular}} & \cellcolor[HTML]{C0C0C0}T-Mobile & AT\&T & Verizon & MetroPCS & \multicolumn{1}{c|}{\begin{tabular}[c]{@{}c@{}}Mint \\ Mobile\end{tabular}} & \cellcolor[HTML]{C0C0C0}Twilio & SlickText & SimpleTexting & text-em-all & \multicolumn{1}{c|}{TextSpot} & Textkiller & Spamhound & Keymessages & Anti-Nuisance & Truecaller & MalwareBytes & Robokiller & NomoRobo & \begin{tabular}[c]{@{}c@{}}Calls \\ Blacklist\end{tabular} & \multicolumn{1}{c|}{\begin{tabular}[c]{@{}c@{}}Call \\ Control\end{tabular}} \\ \midrule
Squatting Techniques & (N=46) & (N=18) & \cellcolor[HTML]{C0C0C0}(N=18) & (N=18) & (N=18) & (N=18) & (N=18) & \cellcolor[HTML]{C0C0C0}N=12 & N=12 & N=12 & N=12 & N=12 & N=12 & N=12 & N=12 & N=12 & N=12 & N=12 & N=12 & N=12 & N=12 & N=12 \\
Typosquatting & 6{[}13.04\%{]} & 3{[}16.67\%{]} & \cellcolor[HTML]{C0C0C0}1{[}33.33\%{]} & 0{[}0.00\%{]} & 0{[}0.00\%{]} & 1{[}33.33\%{]} & 1{[}33.33\%{]} & \cellcolor[HTML]{C0C0C0}0{[}0.00\%{]} & 0{[}0.00\%{]} & 2{[}100\%{]} & 1{[}50.00\%{]} & 0{[}0.00\%{]} & \multicolumn{1}{l}{1{[}50.00\%{]}} & 0{[}0.00\%{]} & 0{[}0.00\%{]} & 0{[}0.00\%{]} & 0{[}0.00\%{]} & 0{[}0.00\%{]} & 2{[}100\%{]} & \multicolumn{1}{l}{1{[}50.00\%{]}} & 0{[}0.00\%{]} & 0{[}0.00\%{]} \\
ComboSquatting & 9{[}19.57\%{]} & 6{[}33.33\%{]} & \cellcolor[HTML]{C0C0C0}2{[}33.33\%{]} & 3{[}50.00\%{]} & 0{[}0.00\%{]} & 2{[}33.33\%{]} & 2{[}33.33\%{]} & \cellcolor[HTML]{C0C0C0}0{[}0.00\%{]} & 0{[}0.00\%{]} & 3{[}75.00\%{]} & 3{[}75.00\%{]} & 0{[}0.00\%{]} & \multicolumn{1}{l}{2{[}50.00\%{]}} & \multicolumn{1}{l}{1{[}25.00\%{]}} & 0{[}0.00\%{]} & 0{[}0.00\%{]} & 0{[}0.00\%{]} & 0{[}0.00\%{]} & \multicolumn{1}{l}{2{[}50.00\%{]}} & \multicolumn{1}{l}{2{[}50.00\%{]}} & 0{[}0.00\%{]} & 0{[}0.00\%{]} \\
Subdomain Usage & 5{[}10.87\%{]} & 4{[}22.22\%{]} & \cellcolor[HTML]{C0C0C0}3{[}75.00\%{]} & 1{[}25.00\%{]} & 0{[}0.00\%{]} & 2{[}50.00\%{]} & 2{[}50.00\%{]} & \cellcolor[HTML]{C0C0C0}0{[}0.00\%{]} & 0{[}0.00\%{]} & 0{[}0.00\%{]} & 0{[}0.00\%{]} & 0{[}0.00\%{]} & \multicolumn{1}{l}{1{[}100\%{]}} & \multicolumn{1}{l}{1{[}100\%{]}} & 0{[}0.00\%{]} & 0{[}0.00\%{]} & 0{[}0.00\%{]} & 0{[}0.00\%{]} & \multicolumn{1}{l}{1{[}100\%{]}} & 0{[}0.00\%{]} & 0{[}0.00\%{]} & 0{[}0.00\%{]} \\
\begin{tabular}[c]{@{}l@{}}Wrong Top-level \\ Domain\end{tabular} & 9{[}19.57\%{]} & 8{[}44.44\%{]} & \cellcolor[HTML]{C0C0C0}4{[}50.00\%{]} & 3{[}37.50\%{]} & 0{[}0.00\%{]} & 3{[}37.50\%{]} & 3{[}37.50\%{]} & \cellcolor[HTML]{C0C0C0}0{[}0.00\%{]} & 0{[}0.00\%{]} & 3{[}75.00\%{]} & 2{[}50.00\%{]} & 0{[}0.00\%{]} & 3{[}75.00\%{]} & 0{[}0.00\%{]} & 0{[}0.00\%{]} & 0{[}0.00\%{]} & 0{[}0.00\%{]} & 0{[}0.00\%{]} & 3{[}75.00\%{]} & 2{[}50.00\%{]} & 0{[}0.00\%{]} & 0{[}0.00\%{]} \\
Homograph & 1{[}2.17\%{]} & 1{[}5.56\%{]} & \cellcolor[HTML]{C0C0C0}0{[}0.00\%{]} & 0{[}0.00\%{]} & 0{[}0.00\%{]} & 0{[}0.00\%{]} & 0{[}0.00\%{]} & \cellcolor[HTML]{C0C0C0}0{[}0.00\%{]} & 0{[}0.00\%{]} & 1{[}100\%{]} & 1{[}100\%{]} & 0{[}0.00\%{]} & 0{[}0.00\%{]} & 1{[}100\%{]} & 0{[}0.00\%{]} & 0{[}0.00\%{]} & 0{[}0.00\%{]} & 0{[}0.00\%{]} & 0{[}0.00\%{]} & 1{[}100\%{]} & 0{[}0.00\%{]} & 0{[}0.00\%{]} \\
None & 28{[}60.87\%{]} & 7{[}38.89\%{]} & \cellcolor[HTML]{C0C0C0}2{[}28.57\%{]} & 2{[}28.57\%{]} & 0{[}0.00\%{]} & 2{[}28.57\%{]} & 2{[}28.57\%{]} & \cellcolor[HTML]{C0C0C0}0{[}0.00\%{]} & 0{[}0.00\%{]} & 1{[}20\%{]} & 1{[}20\%{]} & 0{[}0.00\%{]} & \multicolumn{1}{l}{3{[}60.00\%{]}} & 0{[}0.00\%{]} & 0{[}0.00\%{]} & 0{[}0.00\%{]} & 0{[}0.00\%{]} & 0{[}0.00\%{]} & \multicolumn{1}{l}{3{[}60.00\%{]}} & \multicolumn{1}{l}{2{[}40.00\%{]}} & 0{[}0.00\%{]} & 0{[}0.00\%{]} \\ \midrule
URL Types & (N=46) & (N=18) & \cellcolor[HTML]{C0C0C0}(N=18) & (N=18) & (N=18) & (N=18) & (N=18) & \cellcolor[HTML]{C0C0C0}N=12 & N=12 & N=12 & N=12 & N=12 & N=12 & N=12 & N=12 & N=12 & N=12 & N=12 & N=12 & N=12 & N=12 & N=12 \\
\begin{tabular}[c]{@{}l@{}}Deceptive top-level \\ Domain\end{tabular} & 33{[}71.74\%{]} & 10{[}55.56\%{]} & \cellcolor[HTML]{C0C0C0}1{[}10.00\%{]} & 2{[}20.00\%{]} & 0{[}0.00\%{]} & 1{[}10.00\%{]} & 1{[}10.00\%{]} & \cellcolor[HTML]{C0C0C0}0{[}0.00\%{]} & 0{[}0.00\%{]} & 6{[}66.67\%{]} & 4{[}44.44\%{]} & 0{[}0.00\%{]} & 5{[}55.56\%{]} & 2{[}22.22\%{]} & 0{[}0.00\%{]} & 0{[}0.00\%{]} & 0{[}0.00\%{]} & 0{[}0.00\%{]} & 6{[}66.67\%{]} & 4{[}44.44\%{]} & 0{[}0.00\%{]} & 0{[}0.00\%{]} \\
Unintelligable URL & 9{[}19.57\%{]} & 5{[}27.78\%{]} & \cellcolor[HTML]{C0C0C0}2{[}40.00\%{]} & 2{[}40.00\%{]} & 0{[}0.00\%{]} & 2{[}40.00\%{]} & 2{[}40.00\%{]} & \cellcolor[HTML]{C0C0C0}0{[}0.00\%{]} & 0{[}0.00\%{]} & 0{[}0.00\%{]} & 1{[}33.33\%{]} & 0{[}0.00\%{]} & 2{[}66.67\%{]} & 0{[}0.00\%{]} & 0{[}0.00\%{]} & 0{[}0.00\%{]} & 0{[}0.00\%{]} & 0{[}0.00\%{]} & 2{[}66.67\%{]} & 1{[}33.33\%{]} & 0{[}0.00\%{]} & 0{[}0.00\%{]} \\
\begin{tabular}[c]{@{}l@{}}Long, Deceptive \\ Subdomains\end{tabular} & 3{[}6.52\%{]} & 2{[}11.11\%{]} & \cellcolor[HTML]{C0C0C0}2{[}100\%{]} & 0{[}0.00\%{]} & 0{[}0.00\%{]} & 1{[}50.00\%{]} & 1{[}50.00\%{]} & \cellcolor[HTML]{C0C0C0}- & - & - & - & - & - & - & - & - & - & - & - & - & - & - \\
\begin{tabular}[c]{@{}l@{}}Random domain,\\ deceptive path content\end{tabular} & 1{[}2.17\%{]} & 1{[}5.56\%{]} & \cellcolor[HTML]{C0C0C0}1{[}100\%{]} & 1{[}100\%{]} & 0{[}0.00\%{]} & 1{[}100\%{]} & 1{[}100\%{]} & \cellcolor[HTML]{C0C0C0}- & - & - & - & - & - & - & - & - & - & - & - & - & - & - \\
\begin{tabular}[c]{@{}l@{}}IP address as hostname, \\ deceptive path contents\end{tabular} & 0{[}0.00\%{]} & 0{[}0.00\%{]} & \cellcolor[HTML]{C0C0C0}- & - & - & - & - & \cellcolor[HTML]{C0C0C0}- & - & - & - & - & - & - & - & - & - & - & - & - & - & - \\ \midrule
Named Entities & (N=55) & (N=20) & \cellcolor[HTML]{C0C0C0}N=20 & N=20 & N=20 & N=20 & N=20 & \cellcolor[HTML]{C0C0C0}N=13 & N=13 & N=13 & N=13 & N=13 & N=13 & N=13 & N=13 & N=13 & N=13 & N=13 & N=13 & N=13 & N=13 & N=13 \\
No Entity & 23{[}41.82\%{]} & 5{[}25.00\%{]} & \cellcolor[HTML]{C0C0C0}2{[}40.00\%{]} & 1{[}20.00\%{]} & 0{[}0.00\%{]} & 2{[}40.00\%{]} & 2{[}40.00\%{]} & \cellcolor[HTML]{C0C0C0}0{[}0.00\%{]} & 0{[}0.00\%{]} & 1{[}33.33\%{]} & 1{[}33.33\%{]} & 0{[}0.00\%{]} & \multicolumn{1}{l}{2{[}66.67\%{]}} & 0{[}0.00\%{]} & 0{[}0.00\%{]} & 0{[}0.00\%{]} & 0{[}0.00\%{]} & 0{[}0.00\%{]} & \multicolumn{1}{l}{2{[}66.67\%{]}} & 1{[}33.33\%{]} & 0{[}0.00\%{]} & 0{[}0.00\%{]} \\
Banks & 14{[}25.45\%{]} & 7{[}35.00\%{]} & \cellcolor[HTML]{C0C0C0}4{[}57.14\%{]} & 3{[}42.86\%{]} & 0{[}0.00\%{]} & 3{[}42.86\%{]} & 3{[}42.86\%{]} & \cellcolor[HTML]{C0C0C0}0{[}0.00\%{]} & 0{[}0.00\%{]} & 3{[}100\%{]} & 3{[}100\%{]} & 0{[}0.00\%{]} & 0{[}0.00\%{]} & \multicolumn{1}{l}{1{[}33.33\%{]}} & 0{[}0.00\%{]} & 0{[}0.00\%{]} & 0{[}0.00\%{]} & 0{[}0.00\%{]} & \multicolumn{1}{l}{1{[}33.33\%{]}} & \multicolumn{1}{l}{1{[}33.33\%{]}} & 0{[}0.00\%{]} & 0{[}0.00\%{]} \\
Postal & 6{[}10.91\%{]} & 4{[}20.00\%{]} & \cellcolor[HTML]{C0C0C0}1{[}25.00\%{]} & 1{[}25.00\%{]} & 0{[}0.00\%{]} & 1{[}25.00\%{]} & 1{[}25.00\%{]} & \cellcolor[HTML]{C0C0C0}0{[}0.00\%{]} & 0{[}0.00\%{]} & 1{[}33.33\%{]} & 0{[}0.00\%{]} & 0{[}0.00\%{]} & \multicolumn{1}{l}{3{[}100\%{]}} & 1{[}33.33\%{]} & 0{[}0.00\%{]} & 0{[}0.00\%{]} & 0{[}0.00\%{]} & 0{[}0.00\%{]} & \multicolumn{1}{l}{3{[}100\%{]}} & 1{[}33.33\%{]} & 0{[}0.00\%{]} & 0{[}0.00\%{]} \\
Individual & 4{[}7.27\%{]} & 0{[}0.00\%{]} & \cellcolor[HTML]{C0C0C0}- & - & - & - & - & \cellcolor[HTML]{C0C0C0}- & - & - & - & - & - & - & - & - & - & - & - & - & - & - \\
Social Media & 4{[}7.27\%{]} & 2{[}10.00\%{]} & \cellcolor[HTML]{C0C0C0}0{[}0.00\%{]} & 0{[}0.00\%{]} & 0{[}0.00\%{]} & 0{[}0.00\%{]} & 0{[}0.00\%{]} & \cellcolor[HTML]{C0C0C0}0{[}0.00\%{]} & 0{[}0.00\%{]} & 1{[}50.00\%{]} & 0{[}0.00\%{]} & 0{[}0.00\%{]} & 1{[}50.00\%{]} & 0{[}0.00\%{]} & 0{[}0.00\%{]} & 0{[}0.00\%{]} & 0{[}0.00\%{]} & 0{[}0.00\%{]} & 1{[}50.00\%{]} & 1{[}50.00\%{]} & 0{[}0.00\%{]} & 0{[}0.00\%{]} \\
Other & 4{[}7.27\%{]} & 2{[}10.00\%{]} & \cellcolor[HTML]{C0C0C0}0{[}0.00\%{]} & 0{[}0.00\%{]} & 0{[}0.00\%{]} & 0{[}0.00\%{]} & 0{[}0.00\%{]} & \cellcolor[HTML]{C0C0C0}0{[}0.00\%{]} & 0{[}0.00\%{]} & 1{[}50.00\%{]} & 2{[}100\%{]} & 0{[}0.00\%{]} & 1{[}50.00\%{]} & 0{[}0.00\%{]} & 0{[}0.00\%{]} & 0{[}0.00\%{]} & 0{[}0.00\%{]} & 0{[}0.00\%{]} & 1{[}50.00\%{]} & 1{[}50.00\%{]} & 0{[}0.00\%{]} & 0{[}0.00\%{]} \\ \midrule
Subcategories & (N=55) & (N=20) & \cellcolor[HTML]{C0C0C0}N=20 & N=20 & N=20 & N=20 & N=20 & \cellcolor[HTML]{C0C0C0}N=13 & N=13 & N=13 & N=13 & N=13 & N=13 & N=13 & N=13 & N=13 & N=13 & N=13 & N=13 & N=13 & N=13 & N=13 \\
Account Alert & 14{[}25.45\%{]} & 7{[}35.00\%{]} & \cellcolor[HTML]{C0C0C0}3{[}42.86\%{]} & 3{[}42.86\%{]} & 0{[}0.00\%{]} & 2{[}28.57\%{]} & 2{[}28.57\%{]} & \cellcolor[HTML]{C0C0C0}0{[}0.00\%{]} & 0{[}0.00\%{]} & 3{[}75.00\%{]} & 3{[}75.00\%{]} & 0{[}0.00\%{]} & 0{[}0.00\%{]} & \multicolumn{1}{l}{1{[}25.00\%{]}} & 0{[}0.00\%{]} & 0{[}0.00\%{]} & 0{[}0.00\%{]} & 0{[}0.00\%{]} & \multicolumn{1}{l}{1{[}25.00\%{]}} & \multicolumn{1}{l}{2{[}50.00\%{]}} & 0{[}0.00\%{]} & 0{[}0.00\%{]} \\
Delivery & 9{[}16.36\%{]} & 4{[}20.00\%{]} & \cellcolor[HTML]{C0C0C0}1{[}25.00\%{]} & 1{[}25.00\%{]} & 0{[}0.00\%{]} & 1{[}25.00\%{]} & 1{[}25.00\%{]} & \cellcolor[HTML]{C0C0C0}0{[}0.00\%{]} & 0{[}0.00\%{]} & 1{[}33.33\%{]} & 0{[}0.00\%{]} & 0{[}0.00\%{]} & \multicolumn{1}{l}{3{[}100\%{]}} & 1{[}33.33\%{]} & 0{[}0.00\%{]} & 0{[}0.00\%{]} & 0{[}0.00\%{]} & 0{[}0.00\%{]} & \multicolumn{1}{l}{3{[}100\%{]}} & 1{[}33.33\%{]} & 0{[}0.00\%{]} & 0{[}0.00\%{]} \\
Payday Loan/Credit & 4{[}7.27\%{]} & 2{[}10.00\%{]} & \cellcolor[HTML]{C0C0C0}2{[}100.00\%{]} & 1{[}50.00\%{]} & 0{[}0.00\%{]} & 2{[}100.00\%{]} & 2{[}100.00\%{]} & \cellcolor[HTML]{C0C0C0}- & - & - & - & - & - & - & - & - & - & - & - & - & - & - \\
Finance/Crypto & 2{[}3.64\%{]} & 1{[}5.00\%{]} & \cellcolor[HTML]{C0C0C0}0{[}0.00\%{]} & 0{[}0.00\%{]} & 0{[}0.00\%{]} & 0{[}0.00\%{]} & 0{[}0.00\%{]} & \cellcolor[HTML]{C0C0C0}0{[}0.00\%{]} & 0{[}0.00\%{]} & 1{[}100\%{]} & 0{[}0.00\%{]} & 0{[}0.00\%{]} & 1{[}100\%{]} & 0{[}0.00\%{]} & 0{[}0.00\%{]} & 0{[}0.00\%{]} & 0{[}0.00\%{]} & 0{[}0.00\%{]} & 1{[}100\%{]} & 0{[}0.00\%{]} & 0{[}0.00\%{]} & 0{[}0.00\%{]} \\
Lawsuit/Settlement & 2{[}3.64\%{]} & 0{[}0.00\%{]} & \cellcolor[HTML]{C0C0C0}0{[}0.00\%{]} & 0{[}0.00\%{]} & 0{[}0.00\%{]} & 0{[}0.00\%{]} & 0{[}0.00\%{]} & \cellcolor[HTML]{C0C0C0}0{[}0.00\%{]} & 0{[}0.00\%{]} & 0{[}0.00\%{]} & 0{[}0.00\%{]} & 0{[}0.00\%{]} & 0{[}0.00\%{]} & 0{[}0.00\%{]} & 0{[}0.00\%{]} & 0{[}0.00\%{]} & 0{[}0.00\%{]} & 0{[}0.00\%{]} & 0{[}0.00\%{]} & 0{[}0.00\%{]} & 0{[}0.00\%{]} & 0{[}0.00\%{]} \\
Job Advertisement & 4{[}7.27\%{]} & 2{[}10.00\%{]} & \cellcolor[HTML]{C0C0C0}0{[}0.00\%{]} & 0{[}0.00\%{]} & 0{[}0.00\%{]} & 0{[}0.00\%{]} & 0{[}0.00\%{]} & \cellcolor[HTML]{C0C0C0}0{[}0.00\%{]} & 0{[}0.00\%{]} & 1{[}50.00\%{]} & 1{[}50.00\%{]} & 0{[}0.00\%{]} & \multicolumn{1}{l}{2{[}100\%{]}} & 0{[}0.00\%{]} & 0{[}0.00\%{]} & 0{[}0.00\%{]} & 0{[}0.00\%{]} & 0{[}0.00\%{]} & \multicolumn{1}{l}{2{[}100\%{]}} & \multicolumn{1}{l}{1{[}50.00\%{]}} & 0{[}0.00\%{]} & 0{[}0.00\%{]} \\
Prize/Contest & 12{[}21.82\%{]} & 3{[}15.00\%{]} & \cellcolor[HTML]{C0C0C0}1{[}33.33\%{]} & 0{[}0.00\%{]} & 0{[}0.00\%{]} & 1{[}33.33\%{]} & 1{[}33.33\%{]} & \cellcolor[HTML]{C0C0C0}0{[}0.00\%{]} & 0{[}0.00\%{]} & 1{[}50.00\%{]} & 2{[}100\%{]} & 0{[}0.00\%{]} & 0{[}0.00\%{]} & 0{[}0.00\%{]} & 0{[}0.00\%{]} & 0{[}0.00\%{]} & 0{[}0.00\%{]} & 0{[}0.00\%{]} & 0{[}0.00\%{]} & 0{[}0.00\%{]} & 0{[}0.00\%{]} & 0{[}0.00\%{]} \\
\begin{tabular}[c]{@{}l@{}}Wrong Number/\\ Romance Scam\end{tabular} & 4{[}7.27\%{]} & 0{[}0.00\%{]} & \cellcolor[HTML]{C0C0C0}- & - & - & - & - & \cellcolor[HTML]{C0C0C0}- & - & - & - & - & - & - & - & - & - & - & - & - & - & - \\
Link Only & 3{[}5.45\%{]} & 1{[}5.00\%{]} & \cellcolor[HTML]{C0C0C0}0{[}0.00\%{]} & 0{[}0.00\%{]} & 0{[}0.00\%{]} & 0{[}0.00\%{]} & 0{[}0.00\%{]} & \cellcolor[HTML]{C0C0C0}0{[}0.00\%{]} & 0{[}0.00\%{]} & 0{[}0.00\%{]} & 0{[}0.00\%{]} & 0{[}0.00\%{]} & \multicolumn{1}{l}{1{[}100\%{]}} & 0{[}0.00\%{]} & 0{[}0.00\%{]} & 0{[}0.00\%{]} & 0{[}0.00\%{]} & 0{[}0.00\%{]} & \multicolumn{1}{l}{1{[}100\%{]}} & \multicolumn{1}{l}{1{[}100\%{]}} & 0{[}0.00\%{]} & 0{[}0.00\%{]} \\
Advertisement & 1{[}1.82\%{]} & 0{[}0.00\%{]} & \cellcolor[HTML]{C0C0C0}- & - & - & - & - & \cellcolor[HTML]{C0C0C0}- & - & - & - & - & - & - & - & - & - & - & - & - & - & - \\ \bottomrule
\end{tabular}
}
\caption{The breakdown of the data characterization for all smishing messages selected for the experiment. Highlighted are the Bulk Messenger and Carrier Network used for Test 3. Listed are the number and percentage out of the total possible messages that can be blocked for each Characterization type at the stage in the testing. A '-' means there are no messages left to consider. }
\label{tab:data-characterization}
\end{table}
\end{document}